\documentclass[]{siamart1116}
\usepackage{lipsum}
\usepackage{graphicx}
\usepackage{epstopdf}
\usepackage{algorithmic}
\usepackage{todonotes}
\usepackage{amsmath,amsfonts,bm} 

\usepackage{amsopn}

\ifpdf
  \DeclareGraphicsExtensions{.pdf,.pdf,.png,.jpg}
\else
  \DeclareGraphicsExtensions{.pdf}
\fi
\numberwithin{theorem}{section}

\newcommand{\TheTitle}{Semi-Lagrangian Exponential Integration with application to the rotating shallow water equations } 
\newcommand{\TheAuthors}{P. S. Peixoto, M. Schreiber }

\newcommand{\zoomedareaimage}[1]{\begin{tikzpicture}[remember picture,overlay]
	\node at (-1.9cm,1.7cm) {
		\fcolorbox{red}{red}{\includegraphics[height=2.0cm,trim={5.0cm 4.2cm 11.5cm 3.6cm},clip]{#1}}
	};
\end{tikzpicture}}

\newcommand{\zoomedareaimagex}[1]{\begin{tikzpicture}[remember picture,overlay]
	\node at (3.8cm,2.1cm) {
		\fcolorbox{red}{red}{\includegraphics[height=2.0cm,trim={5.0cm 4.2cm 14.5cm 3.6cm},clip]{#1}}
	};
\end{tikzpicture}}

\newcommand{\zoomedareaimagea}[1]{\begin{tikzpicture}[remember picture,overlay]
	\node at (-2.8cm,1.7cm) {
		\fcolorbox{red}{red}{\includegraphics[height=2.0cm,trim={5.0cm 4.2cm 14.5cm 3.6cm},clip]{#1}}
	};
	\end{tikzpicture}}

\newcommand{\zoomedareaimageax}[1]{\begin{tikzpicture}[remember picture,overlay]
	\node at (11.2cm,2.1cm) {
		\fcolorbox{red}{red}{\includegraphics[height=2.0cm,trim={5.0cm 4.2cm 11.5cm 3.6cm},clip]{#1}}
	};
	\end{tikzpicture}}

\headers{Semi-Lagrangian Exponential Integration}{P. S. Peixoto, M. Schreiber}

\title{{\TheTitle}\thanks{Submitted to the editors 9th of August, 2018.
\funding{This work was funded by FAPESP grants 2014/10501-0 and 2016/18445-7, USP-Santander Mobility grant 527/2016, Shell Brasil under the ANP R\&D levy grant 20714-2 and NCAR funding.}}}

\author{
  Pedro S. Peixoto\thanks{Instituto de Matem\'atica e Estat\'\i stica, Universidade de S\~ao Paulo, Brazil 
    (\email{pedrosp@ime.usp.br}, \url{http://www.ime.usp.br/\~pedrosp}).}
  \and
   Martin Schreiber\thanks{
   	Chair of Computer Architecture and Parallel Systems, Technical University of Munich, Germany (\email{martin.schreiber@tum.de}, \url{http://www.martin-schreiber.info})
	}
}

\ifpdf
\hypersetup{
  pdftitle={\TheTitle},
  pdfauthor={\TheAuthors}
}
\fi

\begin{document}

\maketitle

\begin{abstract}
In this paper we propose a novel way to integrate time-evolving partial differential equations that contain nonlinear advection and stiff linear operators, combining exponential integration techniques and semi-Lagrangian methods. 

The general formulation is built from the solution of an integration factor problem with respect to the problem written with a material derivative, so that the exponential integration scheme naturally incorporates the nonlinear advection. Semi-Lagrangian techniques are used to treat the dependence of the exponential integrator on the flow trajectories. The formulation is general, as many exponential integration techniques could be combined with different semi-Lagrangian methods. This formulation allows an 
accurate solution of the linear stiff operator, a property inherited by the exponential integration technique. It also provides a sufficiently accurate representation of the nonlinear advection, even with large time-step sizes, a property inherited by the semi-Lagrangian method. 

Aiming for application in weather and climate modeling, we discuss possible combinations of well established exponential integration techniques and state-of-the-art semi-Lagrangian methods used operationally in the application. We show experiments for the planar rotating shallow water equations. When compared to traditional exponential integration techniques, the experiments reveal that the coupling with semi-Lagrangian allows stabler integration with larger time-step sizes. From the application perspective, which already uses semi-Lagrangian methods, the exponential treatment could improve the solution of wave-dispersion when compared to semi-implicit schemes.
\end{abstract}



\section{Introduction}


Consider an autonomous initial value problem of the form 
\begin{equation}
	\frac{\partial u}{\partial t}=\mathcal{L}(u)+\mathcal{N}(u), \quad u(0)=u_0
	\label{eq:introgeneralode}
\end{equation}
where $\mathcal{L}$ is a linear (possibly differential) operator and $\mathcal{N}$ is a function (usually nonlinear). 
Exponential integrators are usually derived making use of exponentials of a discrete form of the linear operator $\mathcal{L}$. Many schemes of this form exists, as one may notice from the review of \cite{hochbruck2010exponential}. 


Several application models, such as those related to fluid dynamics \cite{anderson1995computational}, have an important advection term in the equations, usually nonlinear ($\vec{v} \cdot \nabla u$). This can be represented as 
\begin{equation}
\frac{Du}{Dt}=\frac{\partial u}{\partial t}+\vec{v}\cdot \nabla u =\mathcal{L}(u)+\tilde{\mathcal{N}}(u), \quad u(0)=u_0,
\label{eq:generalodeadv}
\end{equation}
where $D/Dt$ represents a total or material derivative, $\vec{v}=\vec{v}(t,\vec{x}, u)$ is the advection velocity, $u=u(t,\vec{x})$, $\tilde{\mathcal{N}}$ represents a general nonlinear term and the gradient ($\nabla$) acts only on the spatial variables ($\nabla=(\partial_{x_1},\partial_{x_2},...,\partial_{x_n}))$. 

The treatment of the nonlinear advection in exponential integrators varies and leads to different mathematical properties.
It can, for instance, be simply thought as a nonlinear term in the exponential integration scheme \cite{beylkin1998new}. Or else, the nonlinear term can be treated via a linearization procedure \cite{Clancy2013, LOFFELD201345,CARR201366,schulze2009exponential, Garcia2014, kooij2018exponential}.

A well-established method to solve equations with nonlinear advection is the semi-Lagrangian advection approach \cite{Robert1981, Staniforth1991, Durran2010}, sometimes denoted as the characteristics method \cite{pironneau1982transport, bermudez1991solving, bermudez2006numerical}. The cost-effectiveness of semi-Lagrangian schemes depends on the problem \cite{bartello1996cost}. They are used in computational fluid dynamics  \cite{XIU2001658, celledoni2016high}, and are very successfully used in weather forecasting \cite{Williamson2007}, hence being adopted by several weather forecasting centres in operational models \cite{diamantakis2013semi,barros1995ifs,figueroa2016brazilian, mengaldo2018current}.


Exponential integrators and semi-Lagrangian schemes have an interesting connection. For linear advection, the characteristics (which define a particle trajectory) are precisely given by the exponential of the linear advection operator \cite{celledoni2005eulerian}. Moreover, for nonlinear advection, it is possible to establish an equivalence between the solution of a general integration factor problem to a semi-Lagrangian approach \cite{st2005nonlinear}. Therefore, it is possible to obtain properties of semi-Lagrangian schemes considering them from an exponential operator point of view.  Or, similarly, it is possible to consider the solution of a semi-Lagrangian problem in place of an operator exponential \cite{celledoni2009semi}. The latter allows, for example, the development of high order semi-Lagrangian schemes \cite{celledoni2016high}.

The goal of this work is to further explore a combination of semi-Lagrangian and exponential integrators. The key development in this paper is to consider an exponential integration scheme that is built with respect to the total (material) derivative, therefore treats nonlinear advection within the exponentiation framework, which, to our knowledge, has not yet been explored in the literature. With this methodology, nonlinear advection is calculated accurately with low dispersion error (property earned from the semi-Lagrangian approach), in combination with an accurate solution of the linear problem even for very stiff problems (property earned from the exponential integration). In principle, several combinations of exponential integration and semi-Lagrangian schemes could be explored. We will derive the general principles of the method and then illustrate how well-established schemes can be used together.

Traditional geophysical fluid dynamics models usually employ either an explicit time stepping scheme, for which the time-step sizes are constrained by faster waves in the system (e.g inertia-gravity), or implicit time stepping schemes (e.g. Crank-Nicolson), which allow larger time-steps, at the cost of slowing the faster (short wavelength) linear waves. 
A recent review on the matter of time stepping schemes for weather and climate \cite{mengaldo2018current} points out the need of time integration schemes that allow large time-steps while preserving wave dispersion properties. Small scale horizontal gravity waves play an important role in the large structure of the middle atmosphere, particularly for climate simulations \cite{mclandress1998importance}. Exponential integrators provide a way to obtain large time-steps without affecting these small-scale waves, preserving superior linear dispersion properties (see 
\cite{Schreiber2018nla,clancy2011laplace}).

An important model for the atmosphere and ocean dynamics is formed by the two-dimensional nonlinear rotating shallow water equations (SWE), as they provide a simple set of equations that already carry many of the complications encountered in full three-dimensional dynamics. Recent works of \cite{Clancy2013} and \cite{GAUDREAULT2016827} explored the use of exponential integrators in SWE and showed its potential and practical relevance to weather forecasting. They explored the dynamic linearization procedure of \cite{tokman2006efficient} to obtain their exponential integrator, and the nonlinear advection was treated within the linearization. Also within this application framework, \cite{Garcia2014} shows results from exponential integrator schemes for Boussinesq thermal convection, indicating higher computational cost but greater accuracy with respect to well established schemes for the problem. Considering linear equation sets for this application, \cite{archibald2009time} solves the linear advection problem on the sphere, which is an important test case for weather and climate, using exponential integration.
Also, \cite{schreiber2017beyond} solves the linear SWE on the plane and spectral solver with a rational approximation of exponential integrator \cite{Haut2015} (T-REXI) and analyze the potential computational gain of a massively parallel scheme to compute exponentials on linear operators.
The full non-linear SWE on the rotating sphere is solved in \cite{schreiber_2019_exp_nonlinearswe_sphere} with a numerical Cauchy-contour-integral approximating exponential time integration (CI-REXI), resulting in promising wallclocktime-to-error improvements with exponential integrators.
However, the practical adequacy of exponential integration schemes with semi-Lagrangian methods for weather and climate is still a matter of research. In particular, it turns out that taking large time-steps with semi-Lagrangian formulations is extremely challenging, for which this study hopes to contribute.

A combination with similarities to the one proposed here was developed by \cite{clancy2011laplace} where, instead of deriving the exponential integration along trajectories, a Laplace transform following trajectories was used. They analyze how this semi-Lagrangian Laplace transform method can improve certain aspects of the solutions obtained with traditional semi-Lagrangian semi-implicit scheme considering also a shallow water model.
However, obtaining large time-step sizes turned out be a challenge in their formulation.

The paper is organized as follows. In Section \ref{sec:expint} and \ref{sec:sl} we review usual exponential integration techniques and semi-Lagrangian techniques, respectively. These two sections will be used in the development of the novel semi-Lagrangian exponential technique, which is shown in Section \ref{sec:slexp}. Section \ref{sec:swe} reviews properties of the SWE. Numerical results of the methods developed are shown in Section \ref{sec:num}. We finish the paper with some remarks in Section \ref{sec:conc}.



%
%
%

\section{Exponential integration}
\label{sec:expint}
We start providing a brief review of some existing exponential integration techniques that will be relevant for the semi-Lagrangian exponential approach. More details may be found in the review of exponential integrators of \cite{hochbruck2010exponential} and in references therein.

\subsection{Analytical time integration}

Numerically, the solution of equation \eqref{eq:introgeneralode}, $u(t)$, is approximated by ($n$) discrete values that could be, for example, grid point values or spectral coefficients. This defines the discrete solution $U(t) \in \mathbb{R}^n$ evolving in time. The linear operator ($\mathcal{L}$) can be approximated by a discrete version of it ($L$), with a preferred discretization scheme. Since $L$ may be originated from a partial differential equation problem, it is prudent to keep in mind that $L$ may carry information from spatially varying features.
However, having derived it for an autonomous system, it is independent of time. So the analogous semi-discrete problem of interest may be written as
\begin{equation}
\frac{dU(t)}{dt}=LU(t)+N(U(t)), \quad U(0)=U_0,
\label{eq:generaldiscreteode}
\end{equation}
where $L\in \mathbb{R}^n\times \mathbb{R}^n$ is the discrete linear operator (an $n\times n$ matrix) and $N(U)$ is a discrete version of $\mathcal{N}(u)$.


Now let's assume that $U(t_n)$ is given for a current time $t_n$, and that we wish to calculate $U(t_{n+1})$, for $t_{n+1}=t_n+\Delta t$. Since $L$ does not depend on time, the integration factor problem,
\begin{equation}
\frac{dQ_n(t)}{dt}=-Q_n(t)L, \quad Q_n(t_n)=I,
\label{eq:intfact}
\end{equation}
where $I$ is the identity matrix, has a unique solution given by
\begin{equation}
Q_n(t)=e^{-(t-t_n)L}.
\label{eq:Qn}
\end{equation}
Using the integration factor in equation \eqref{eq:generaldiscreteode} one sees that 
\begin{equation}
\frac{d }{dt}\left(Q_n(t)U(t)\right)=Q_n(t)N(U).
\label{eq:generaleqexp}
\end{equation}
Therefore the problem has an exact solution which may be implicitly represented as,
\begin{equation}
U(t_{n+1})=e^{\Delta t L}U(t_n)+e^{\Delta t L}\int_{t_n}^{t_{n+1}} e^{-(s-t_{n})L}N(U(s)) ds,
\label{eq:generalexpsol}
\end{equation}
where we note that $Q^{-1}_n(t)=e^{(t-t_n)L}$ is the inverse of $Q_n(t)$. This last equation is well-known as the variation-of-constants formula. 

\subsection{Numerical time integration (ETDRK)}

Exponential integration makes use of calculations of the exponentials, and/or exponential related functions, to obtain a time marching scheme along the lines of equation \eqref{eq:generalexpsol}.

The main difference between variants of 
exponential integration schemes is in the way the nonlinear term is evaluated.
If the equation is purely linear ($N=0$), then the integral term in equation \eqref{eq:generalexpsol} vanishes and
it is possible to solve the problem directly from the matrix exponential calculation for each time-step. For nonlinear problems, there exists several approaches (see \cite{hochbruck2010exponential}). We will use the Runge-Kutta Exponential Time Differencing (ETDRK) as a representative method, following \cite{cox2002exponential}. However, for the semi-Lagrangian exponential scheme (to be shown), other methods could be considered in a similar fashion. 

As a first order approximation, let the nonlinear term $N(U)$ in the integral be constant in time, for each time-step, with value $N(U(t_n))$. Using equation \eqref{eq:intfact} and assuming $L^{-1}$ exists, we may then formally derive what is known as the first order ETD1RK method,
\begin{equation}
{U}_\text{ED1}^{n+1}= \varphi_0(\Delta t L)U^n+\Delta t \, \varphi_1(\Delta t L) N(U^n)  ,
\end{equation}
where, 
\begin{equation}
\varphi_0(z)=e^{z}, \quad \quad  \varphi_1(z)=z^{-1}(e^{z}-1).
\label{eq:phi1}
\end{equation}
with $z=\Delta t L$.
%

More general (higher order) ETD schemes may be derived using higher order $\varphi_k$ functions (see \cite{cox2002exponential}), which may be defined from the recurrence relation
\begin{equation}
	\label{eq:varphi_n}
	\varphi_{k+1}(z)=z^{-1}\left(\varphi_k(z)-\varphi_k(0)\right), 
\end{equation}
where potential singularities may be treated by a series expansions for $\varphi_n$ evaluations close to their singularity. 

We will be particularly interested in this paper in the second order ETD2RK scheme, in order to allow a fair comparison to other well-established second order approaches in our numerical experiments. Let $U^n$ be the numerical approximation of $U(t_n)$ at time $t$, then the ETD2RK scheme may be written, as function of the ETD1RK scheme, as

\begin{equation}
U^{n+1}_\text{ED2}=  {U}_\text{ED1}^{n+1} + \Delta t \varphi_2(\Delta t L)\left( N({U}_\text{ED1}^{n+1})-N(U^n) \right),
\label{eq:etd2}
\end{equation}
which is derived substituting the second order approximation for the nonlinear term,
\begin{equation}
N(U(s))=N(U(t_n))+\frac{(s-t_n)}{\Delta t}\left(N({U}_\text{ED1}(t_{n+1}))-N(U(t_n))\right)+\mathcal{O}(\Delta t^2),
\end{equation}
into equation \eqref{eq:generalexpsol}.

%
\section{Semi-Lagrangian integration}
\label{sec:sl}

Eulerian schemes usually keep a fixed grid and evaluate the movement of the particles that pass through a computational cell. For nonlinear advection, these schemes usually have time-step size limited by the Courant-Friedrichs-Lewy condition (CFL).
Lagrangian schemes usually follow particle trajectories (characteristics) through time and may not even rely on a fixed computational grid, or else have a grid evolving over time. This can create complicated grid structures involving, for example, intersections of trajectories.
Semi-Lagrangian schemes compensate for this by keeping a fixed grid, but following the particle trajectories for a single time-step (a local version of the classical Lagrangian approach). Since the trajectories may end, or start, in points not on the reference grid, usually an interpolation step is required.

In the context of atmospheric simulations, this scheme usually allows time-step sizes significantly larger than CFL-restricted Eulerian schemes \cite{robert1982semi}. 
As we will observe, maintaining this property of significantly larger time-step sizes is a non-trivial task with exponential integration.

In this section we introduce classic notations and results about semi-Lagrangian schemes. This will be required as a basis to derive the semi-Lagrangian exponential schemes in the next section.
Further details on semi-Lagrangian methods can be found in \cite{Staniforth1991} and \cite{Durran2010}.

\subsection{The material derivative}



We start considering Equation \eqref{eq:generalodeadv} on a Lagrangian framework, relative to a particle initially positioned at $\vec{r}_0$ in space. Thus, the system state is formed by $u=u(t,\vec{r}(t))$, with advection velocity defined as $\vec{v}=\vec{v}(t,\vec{r}(t), u(t, \vec{r}(t)))$. Here, $\vec{r}(t)$ is the Lagrangian trajectory of the particle, therefore it is the solution of the non-autonomous problem 
\begin{equation}
\frac{d \vec{r}(t)}{dt}=\vec{v}(t, \vec{r}(t), u(t, \vec{r}(t))), \quad \vec{r}(0)=\vec{r}_0.
\label{eq:traj}
\end{equation}
Equation \eqref{eq:generalodeadv} may be written in a Lagrangian framework as
\begin{equation}
\frac{d u(t,\vec{r}(t))}{d t} =\frac{\partial u(t,\vec{r}(t))}{\partial t}+\vec{v}\cdot \nabla u(t,\vec{r}(t))= \mathcal{L}(u(t,\vec{r}(t)))+\tilde{\mathcal{N}}(u(t,\vec{r}(t))), 
\label{eq:generalodelag}
\end{equation}
with initial condition $u(0, \vec{r}_0)=u_0$, where now $\mathcal{L}$ and $\tilde{\mathcal{N}}$ may implicitly also depend on the position $\vec{r}(t)$.
This particular time derivative ($d/dt$) on the Lagrangian framework is usually denoted as a total (material) derivative $D/Dt$, as in equation \eqref{eq:generalodeadv}.

As in the previous section, we will focus here on a discretized problem, where $\mathcal{L}$ may be again directly viewed as a finite dimensional matrix operator, hence linear, and will be denoted by $L$. In a Lagrangian framework, $L$ may depend on the particle position $\vec{r}(t)$. Therefore, we may analogously to equation \eqref{eq:generaldiscreteode} set the general non-autonomous semi-discrete problem to be
\begin{equation}
\frac{DU(t,\vec{r}(t))}{Dt}=L(U(t,\vec{r}(t)))+\tilde{N}(U(t,\vec{r}(t))), \quad U(t_0,\vec{r}(t_0))=U^0,
\label{eq:generalodelagdisc}
\end{equation}
where $\tilde{N}$ is a numerical approximation to $\tilde{\mathcal{N}}$, and now the time differential is a total derivative and depends on the solution of the problem given in \eqref{eq:traj}.

Although there are many forms of semi-Lagrangian schemes \cite{Staniforth1991}, these usually rely on basically two components with both playing important roles in the accuracy and stability of the schemes \cite{falcone1998convergence,Durran2010, Peixoto2014}.

Component (i): interpolation of the information to the reference grid. As shown in \cite{falcone1998convergence}, the interpolation order needs to be chosen in agreement with the accuracy order of the trajectory calculation.

Component (ii): the evaluation of trajectories, which are solutions of the problem \eqref{eq:traj}.

We will consider a back-trajectory approach, which is a well-established approach \cite{Hortal2002} that assumes that the grid is fixed at time $t_{n+1}$. The trajectory determines the position of a departure point at time $t_n$, which is likely not to be a grid point, so an interpolation of the advected quantity is required.
The trajectory evaluation itself can then be obtained by a direct numerical time integration of differential equation \eqref{eq:traj}, as a sub-cycling procedure, or, which is more common in atmospheric applications, iteratively solve its integral form. In the later,
one may obtain the departure point $\vec{r}_d=\vec{r}(t_{n})$ from the knowledge of the arrival point $\vec{r}_{a}=\vec{r}(t_{n+1})$, which is set to be a grid point, using two-time level schemes \cite{McDonald1987}. This can be done using the midpoint rule integration (for  $\vec{r}_m=\vec{r}(t_{n+1/2})$) and an iterative procedure to solve the nonlinear resulting equation. An example of such procedure, which will be used in the present work,
uses the following iterative equation,
\begin{equation}
\vec{r}^{\,k+1}_m = \vec{r}_a-\vec{v}(t_{n+1/2}, \vec{r}^{\,k}_m, u^k_m) \frac{\Delta t}{2},
\end{equation}
where  $u^k_m=u(t_{n+1/2},\vec{r}^{\,k}_m)$ and usually  $\vec{r}^{\,0}_m=\vec{r}_a$ is used as initial step for the procedure. The departure point is then obtained by simply considering $\vec{r}_d=2\vec{r}_m-\vec{r}_a$. 

In case $\vec{v}$ is not known within $[t_n,t_{n+1}]$, for example
if $\vec{v}$ depends on $u$, its evaluation in intermediate times requires an extrapolation from previous time-steps. This extrapolation may directly influence the stability of the scheme \cite{Durran2010}. A well-established approach is the Stable Extrapolation Two-Time-Level Scheme (SETTLS) of \cite{Hortal2002}, used at the ECMWF\footnote{European Centre for Medium-Range Weather Forecasts} in their global spectral model IFS for operational weather forecasts. All trajectory calculations used in this work follow the SETTLS scheme to obtain the departure points.

\subsection{Semi-Lagrangian Solver (SL-SI-SETTLS)}
\label{sec:sl-settls}
An important scheme for atmospheric modeling is the one used in the IFS-ECMWF model. It uses a semi-Lagrangian scheme coupled with a semi-implicit time stepping of linear terms with spectral horizontal discretization. This scheme, based on \cite{Hortal2002}, will serve as a first guideline in the development of the semi-Lagrangian exponential schemes.

The semi-implicit discretization with semi-Lagrangian Crank-Nicolson time stepping assumes
\begin{equation}
\frac{U^{n+1}-U^{n}_*}{\Delta t} = \frac{1}{2}\left( (LU)^{n+1}+(LU)^n_*\right)+\tilde{N}^{n+1/2},
\label{eq:ifsscheme}
\end{equation}
where the subscript $_*$ with superscript $^n$  denotes interpolation to departure points ($\vec{r}_d^{}$) \cite{bates1990integration} and the last term represents the non-linearities at the midpoint of the trajectory. This term is computed based on averaging and extrapolation (see \cite{Hortal2002}, Eq.\,(4.4,4.5)) with 
\begin{equation}
	\tilde{N}^{n+1/2} = \frac{1}{2}\Big(\underbrace{\left[2\tilde{N}^n-\tilde{N}^{n-1}\right]_*}_{\text{Extrapolation to $t_{n+1}$}} + \tilde{N}^n \Big),
\label{eq:ifsscheme2}
\end{equation}
which is the SETTLS extrapolation, where $\tilde{N}^n$ is the evaluation of the nonlinear term at time $t_{n}$. The unknowns in Equation \eqref{eq:ifsscheme} are implicitly given by $U^{n+1}$ and $(LU)^{n+1}$, requiring a linear solver. 

 
To ensure an overall second order accurate scheme (assuming $\Delta t \propto \Delta x$), it is sufficient to use cubic interpolations of the advected quantities (with respect to Equations \eqref{eq:ifsscheme} and \eqref{eq:ifsscheme2}), and linear interpolations of the velocities in the iterative process of trajectory calculations \cite{Peixoto2014}.

\section{Semi-Lagrangian exponential integration}
\label{sec:slexp}

In this section, we discuss how the general exponential integration techniques can be applied in a Lagrangian reference frame, to derive the novel semi-Lagrangian exponential methodology.

\subsection{Basic theory}


The key concept investigated in this paper is to consider, from a numerical perspective, the exponential integration of Equation \eqref{eq:generalodelagdisc} considering the total (material) derivative.

As in Section \ref{sec:expint}, where we built exponential integration schemes from the solution of an integration factor problem, we would like to be able to define a similar integration factor for the problem with respect to this material derivative.
We assume the existence of an integration factor $P_n(t)$ that is a solution to the problem 
\begin{equation}
\frac{D (P_n(t)\,U(t, \vec{r}(t)))}{Dt}=P_n(t)\tilde{N}(U(t,\vec{r}(t))), \quad P_n(t_n)=I.
\label{eq:lagintfact}
\end{equation}
Assuming  that $U$ is a solution of \eqref{eq:generalodelagdisc}, $P_n$ will also be a solution of
\begin{equation}
\frac{D P_n(t)}{Dt}U(t,\vec{r}(t))=-P_n(t)L(U(t,\vec{r}(t))), \quad P_n(t_n)=I.
\label{eq:lagintfactor}
\end{equation}
We recall that $L$ may depend on the spatial variations, which are now depending also on time due to the Lagrangian framework, so we will explicitly indicate this with a subscript as $L=L_{\vec{r}(t)}$. If $L_{\vec{r}(t)}$ commutes in time, that is, $L_{\vec{r}(t)}L_{\vec{r}(s)}=L_{\vec{r}(s)}L_{\vec{r}(t)}$ for all times $s$ and $t$, then the integration factor problem has a solution given by
\begin{equation}
P_n(t)=e^{-\int_{t_n}^{t}L_{\vec{r}(s)}ds}.
\end{equation}
For the continuous problem with $\mathcal{L}$ based on space-varying coefficients (dependent of the particle position), the commutation assumption of $L$ will most likely not be satisfied. The integration factor may, however, still exist and be well defined, but might not have the usual matrix exponential form. Assuming that such an integration factor exists, and that it is invertible ($P_n^{-1}$ exists for all time), equations \eqref{eq:generalodelagdisc} and \eqref{eq:lagintfact} indicate the following implicit relation on $U$ (analogous to \eqref{eq:generalexpsol}),
\begin{equation}
U(t_{n+1}, \vec{r}(t_{n+1}))=P^{-1}_n(t_{n+1})U(t_{n}, \vec{r}(t_{n}))+P^{-1}_n(t_{n+1}) \int_{t_n}^{t_{n+1}} P_n(s)\tilde{N}(U(s, \vec{r}(s))) ds.
\label{eq:generaleqexpP}
\end{equation}
This is the fundamental equation for the derivation of the semi-Lagrangian exponential schemes developed in this paper.

Numerically, one needs an explicit way of calculating the integration factor. This will depend on the problem of interest. One possibility is to directly numerically integrate equation \eqref{eq:lagintfactor}, which is the basis of many operator splitting techniques \cite{st2005nonlinear}. 

Another possibility, if such integration factor is unknown in its exponential form, is to assume  that $L$ does not vary within each time-step for each given local trajectory, since then the problem reduces to a matrix exponential problem. This assumption can be also thought as the fundamental idea of semi-Langrangian methods which we will apply later: interpolating the Lagrangian solution to a regular Eulerian grid allows applying the linear and nonlinear operators on a regular grid. 
This should provide a first order approximation to the true integration factor at each time-step. This greatly simplifies the problem, as in this case $P_n=Q_n$, as defined in equation \eqref{eq:Qn}, and the problem reduces to
\begin{equation}
U(t_{n+1}, \vec{r}(t_{n+1})) = e^{\Delta t L}U(t_{n}, \vec{r}(t_{n}))+e^{\Delta t L}\int_{t_n}^{t_{n+1}}e^{-(s-t_n)L} \tilde{N}(U(s, \vec{r}(s))) ds + E_L,
\label{eq:generaleqexpPconst}
\end{equation}
where $E_L$ denotes the potential errors introduced with the assumption of $L$ being constant along the trajectories.
This is almost identical to what we obtained for the usual exponential integration approach (see equation \eqref{eq:generalexpsol}), but now $U$ is varying along a particle trajectory in time, resulting in a derivation of what we are calling a semi-Lagrangian exponential integration.
Using the semi-Lagrangian notation, we rewrite the numerical method from
equation \eqref{eq:generaleqexpPconst} as
\begin{equation}
	U^{n+1}=e^{\Delta t L}U^n_*+e^{\Delta t L}\int_{t_n}^{t_{n+1}}e^{-(s-t_n)L} \tilde{N}(U(s, \vec{r}(s))) ds + E_L,
	\label{eq:generaleqexpL}
\end{equation}
where $U^{n+1}$ is given at grid points and $U^n_*$ refers to the (interpolated) value at departure points. 


Finally, different semi-Lagrangian exponential schemes can be built depending on how the integral is approximated, as is the case with the usual exponential integration techniques. The integral consists of an application of a linear operator, which we will compactly write as $T(s)=e^{-(s-t_n)L}$, and a nonlinear function calculation, compactly $w(s)=\tilde{N}(U(s, \vec{r}(s)))$. Interestingly, a method as simple as calculating a midpoint rule integration can have different forms, and naive formulations may easily generate inconsistent schemes. 

We will discuss this issue by first pointing out two very important remarks which will play a crucial role for the derivation of a consistent and stable semi-Lagrangian exponential integration scheme:
\begin{enumerate}
	\item[(R1)] \textit{Integration}: The integral term relies on a linear operator, $T$, acting on a nonlinear function, $w$, integrated along a trajectory.
	If we wish to evaluate the term $Tw$ at time $t_{n-k}$, with $k\geq0$, at departure points (or trajectory midpoints), we should first apply the linear operator to the nonlinear function evaluated at grid points at time $t_{n-k}$, and only then interpolate to the desired trajectory point.
	Otherwise, if we first interpolate the nonlinear function to the departure points, then the application of the linear operator would be referring to an irregular grid, therefore possibly not being well defined numerically.
	Consequently, \emph{at time $t_{n-k}$, the application of the linear operators should come {before} the interpolation operation}, as, for example, with $\left( T(t_{n-k})w(t_{n-k}) \right)_*$.
	
	\item[(R2)] \textit{Advected quantities}: At time $t_{n+1}$, interpolated values of quantities coming from times $t_{n-k}$, $k\geq0$, are assumed to have already been advected, therefore the results lay on a regular grid relative to the arrival points.
	Consequently, \emph{at time $t_{n+1}$, for advected quantities, the linear operator should be applied after the interpolation operation}, as, for example, with  $e^{\Delta t L}U_*^n$ term of equation \eqref{eq:generaleqexpL}.
\end{enumerate}


These important remarks are a peculiarity of this kind of semi-Lagrangian exponential formulation. They are required since the interpolation operation of advected quantities in general does not commute with a linear operator. 
That is, even thought $e^{\Delta t L}$ is a linear operator, possibly independent of time and space, it does not in general commute with the interpolation operator $(_*)$, since this interpolation reflects a non-regular grid formed by nonlinear backward trajectories. Therefore, in general, $e^{\Delta tL} U^{n}_{*} \neq \left( e^{\Delta tL}  U^{n}\right)_{*}$. We provide, 
in Appendix \ref{ap:slexpprop},
an illustration of this lack of commutation, which interestingly happens even in the case of linear advection.

Returning to the case of the midpoint rule integration of the nonlinear term, we may have, for example, these 2 distinct approximations of the integral term,
\begin{equation}
 \int_{t_n}^{t_{n+1}} T(s) w(s)\, ds \approx \begin{cases}
A_1 = & \Delta t \, T(t_{n+1/2})\left[w(t_{n+1/2})\right]_\dagger \\
A_2 = & \Delta t \, \left[T(t_{n+1/2})w(t_{n+1/2})\right]_\dagger
                                             \end{cases},
\end{equation}
where we used the $\dagger$ symbol to indicate the interpolation to the midpoints of the trajectories (whereas the interpolation to departure points are indicated with $*$). 

In $A_1$, first the nonlinear function ($w$) is evaluated at the intermediate time ($t_{n+1/2}$) on a regular grid and then this is used to obtain  interpolated values at the trajectories midpoints. Only then the operator $T$, evaluated at time $t_{n+1/2}$, is applied. Since the operator is applied on values given at trajectories midpoints, this application happens on a possibly irregular grid, therefore may results in an inconsistent application of the operator.

In $A_2$, first both the operator $T$ and the nonlinear function $w$ are evaluated at the intermediate time ($t_{n+1/2}$), on a regular grid, then the operator is applied on $w$. Only then  interpolated values are obtained for the midpoints of the trajectories. In this case the linear operator is always applied on regular grid quantities.

As an example, consider the following scheme,
\begin{equation}
U^{n+1} = e^{\Delta t L} U_*^n + \Delta t\, e^{\frac{\Delta t}{2} L}\tilde{N}^{n+ 1/2}, \quad \text{(unstable scheme)}.
\label{eq:slexp-unstable}
\end{equation}
where $\tilde{N}^{n+1/2}$ is an approximation of the nonlinear term at the trajectory midpoint, for example considering equation \eqref{eq:ifsscheme2}.
Due to the aforementioned remarks and discussion, this scheme applies the linear operator on conceptually irregular grids, and therefore may be inconsistent with the underlying equations. Experiments with this method revealed it is unstable even with small time-step sizes.

%
%
%
%
%

\subsection{Semi-Lagrangian Exponential SETTLS (SL-EXP-SETTLS)}
\label{sec:sl-exp-settls}
Following the SETTLS scheme \cite{Hortal2002} for the semi-Lagrangian discretization, but using it with respect to equation \eqref{eq:generaleqexpL}, we may derive our first combination of semi-Lagrangian exponential scheme, which we will denote as SL-EXP-SETTLS. The scheme is derived from \eqref{eq:generaleqexpL} as
\begin{equation}
U^{n+1}_\text{SLEX}=e^{\Delta tL}  U^{n}_{*}+\Delta t \, e^{\Delta tL}  \tilde{N}_e^{n+1/2},
\label{eq:sl-settls}
\end{equation}
where we use the SETTLS extrapolation to obtain the value of $\tilde{N}$ at the trajectory midpoint as
\begin{equation}
 \tilde{N}_e^{n+1/2} = 
\frac{1}{2}\left[2 \tilde{N}^{n} - e^{\Delta tL} \tilde{N}^{n-1} \right]^n_*+  \frac{1}{2}\tilde{N}^{n}.
\label{eq:sl-settls-extrap}
\end{equation}
We note that $ \tilde{N}_e^{n+1/2} $ is an approximation of $e^{-(s-t_n)L} \tilde{N}(U(s, \vec{r}(s)))$ at the midpoint of the trajectory taking into account the remark (R1).
One may note that it differs from the method presented in Equation \eqref{eq:slexp-unstable}, as in this case the linear operator and nonlinear calculations are not split, but computed jointly at desired time-steps. 
%

It is possible to simplify the above equations in order to require only 2 exponential evaluations per time-step. This scheme is a multi-step scheme, that requires information from two previous time-steps.
This scheme may also be thought as a semi-Lagrangian version of the Integrating Factor method, proposed in \cite{cox2002exponential} as the second order Adams-Bashforth Integrating Factor method (IFAB2), as one can notice from their equation (31). 

As discussed in \cite{cox2002exponential}, the concept of stability for Integrating Factor methods is unclear. This is also the case for our semi-Lagrangian version of exponential schemes. Therefore, this is a topic we discuss in this paper purely from a numerical perspective.


\subsection{Semi-Lagrangian Exponential ETDRK (SL-ETDRK)}
\label{sec:sl-etdrk}
To construct semi-Lagrangian Exponential Time Differencing Runge-Kutta schemes (SL-ETDRK) in analogy to usual ETDRK schemes, we need to pay attention to the remarks (R1) and (R2) above. In usual ETD schemes, as shown in Section \ref{sec:expint}, the exponential in front of the integral in equation \eqref{eq:generaleqexpL} would commute with the integral, to be placed  within the integrand. However, since now the integral is along trajectories, this no longer results in an equivalent problem in the numerical scheme, due the remarks pointed out above. Therefore, we should first evaluate the integral term, and then apply the linear operator ($e^{\Delta t L}$).

Following this strategy, we may derive the semi-Lagrangian ETD1RK scheme in the following way. From equation \eqref{eq:generaleqexpPconst}, assuming as in ETD1RK that the non-linearity is constant within a time-step, we have
\begin{equation}
{U}_\text{SLED1}^{n+1}= \varphi_0(\Delta t L) \left[ U^n + \Delta t \, \varphi_1(-\Delta t L) \tilde{N}(U^n)\right]_*^n,
\end{equation}
where we used that $\varphi_1(z)=\varphi_0(z)\varphi_1(-z)$. This scheme can be computed numerically with only two $\varphi$ function evaluations and one interpolation per time-step.

Deriving the second order scheme (SL-ETD2RK) involves a more careful analysis of how the integral in equation \eqref{eq:generaleqexpPconst} is approximated. Let
\begin{eqnarray}
\nonumber N(U(s)) & = & N(U(t_n,\vec{r}(t_n)))+ \\ 
 &&\frac{(s-t_n)}{\Delta t}\left(N(U_\text{SLED1}(t_{n+1}, \vec{r}(t_{n+1})))-N(U(t_n, \vec{r}(t_n)))\right)+\mathcal{O}(\Delta t^2),
\end{eqnarray}
then 
 \begin{eqnarray}
\nonumber U_\text{SLED2}(t_{n+1}, \vec{r}(t_{n+1}))&=&  \varphi_0(\Delta t L)U(t_{n}, \vec{r}(t_{n}))\\
\nonumber &+& \varphi_0(\Delta t L)\left( \int_{t_n}^{t_{n+1}}e^{-(s-t_n)L} ds \right) N(U(t_n, \vec{r}(t_n)))  \\
\nonumber &+& \varphi_0(\Delta t L)\left( \int_{t_n}^{t_{n+1}}\frac{(s-t_n)}{\Delta t} e^{-(s-t_n)L}ds\right) N(U(t_{n+1}, \vec{r}(t_{n+1}))) \\
&-& \varphi_0(\Delta t L)\left( \int_{t_n}^{t_{n+1}}\frac{(s-t_n)}{\Delta t} e^{-(s-t_n)L} ds\right) N(U(t_n, \vec{r}(t_n))) .
\end{eqnarray}

To be able to preserve $e^{\Delta t L}$ outside of the integral, and still make use of the $\varphi$ functions of ETDRK schemes, we may factor out the $\varphi_0(z)=e^z$ function of the $\varphi_k$ functions. Let $\psi_k$ functions be defined as
\begin{equation}
	\psi_k(z)= (-1)^{k+1} \varphi_k(-z) + \sum_{l=1}^{k-1}\varphi_l(-z).
\end{equation}
It can be shown that $\varphi_k(z)=\varphi_0(z)\psi_k(z)$ by substituting equation \eqref{eq:varphi_n} into the right-hand-side of the definition of $\psi_k$ and using binomial expansions in a similar way as done in \cite{cox2002exponential}.

Using the SL-ETD1RK scheme and the properties of the $\varphi$ functions with respect to $\psi$ functions, we may write the SL-ETD2RK scheme as
\begin{equation}
U^{n+1}_\text{SLED2}=  {U}_\text{SLED1}^{n+1} + \Delta t\, \varphi_0(\Delta t L)\left[ \psi_2(\Delta t L) N({U}_\text{SLED1}^{n+1}) - \left(\psi_2(\Delta t L)N(U^n) \right)_*^n\right].
\end{equation}


After suitably rearranging the equations, the scheme can be coded to require 4 evaluations of $\varphi$ (or $\psi$) functions and 2 interpolations.

%

\section{Rotating Shallow Water Equations on f-Plane}
\label{sec:swe}
%


In this section we review the basic concepts of the Shallow Water Equations (SWE), which will serve as an application for the schemes discussed in the previous sections.

Considering a Lagrangian framework, with particle trajectories given by $\vec{r}(t)=(x(t), y(t))$ on a plane, we define $\vec{v}=\vec{v}(t,\vec{r}(t))=(u(t,\vec{r}(t)), v(t,\vec{r}(t)))$ to be the flow velocity, and $\eta=\eta(t,\vec{r}(t))$ a fluid depth perturbation about a constant mean fluid depth ($\bar{\eta}$). The rotating SWE on a plane may then be written as
 \begin{equation}
\frac{DU}{Dt}=\mathcal{L}U+\tilde{\mathcal{N}}(U), 
 \end{equation}
 where the time derivative is assumed to be the total (material) derivative, and
  \begin{equation}
  \label{eq:swe}
U=\left(\begin{array}{c}
  u\\
 v\\
 \eta  
\end{array}\right),\quad
\mathcal{L}=
\left(\begin{array}{ccc}
  0 & f & -g\partial_{x}\\
 -f & 0 & -g\partial_{y}\\
 -\bar{\eta} \partial_{x}& -\bar{\eta} \partial_{
 y
 } & 0  
\end{array}\right),
\quad 
 \tilde{\mathcal{N}}(U)=\left(\begin{array}{c}
  0\\
 0\\
 -\eta \nabla\cdot \vec{v}  
\end{array}\right),
 \end{equation}
where the total fluid depth $h$ is given by $h=\eta+\bar{\eta}$. The gravity $g$ and the Coriolis parameter $f$ are  assumed to be constant throughout this paper (f-plane approximation). Initial conditions for the prognostic variables $(u, v, \eta )$ are assumed to be given. Bi-periodic boundary conditions will be adopted for $(x,y)$ on a rectangular limited set of $\mathbb{R}^2$. The bottom topography considered in this work is assumed to be flat.

Well-established  models adopt semi-implicit  schemes \cite{Durran2010, robert1982semi}, with implicit treatment of linear terms and explicit treatment of non-linearities. Among the implicit schemes for the linear waves, Crank-Nicolson (trapezoidal differencing) is frequently adopted, as done for example in the IFS model of the ECMWF \cite{ifs,Hortal2002}, coupled with a semi-Lagrangian approach. Modern models that use non-regular spherical grids, such as MPAS \cite{Skamarock2012} or DYNAMICO \cite{Dubos2015}, adopt explicit time stepping procedures based on Runge-Kutta time integration. See \cite{mengaldo2018current} for an extensive list and description of the main time stepping schemes used for weather and climate models.



 \subsection{Exponential of the linear operator}

We seek to find the exponential of the linear operator $\mathcal{L}$ where we assume the time-step size $\Delta t$ incorporated into $\mathcal{L}$ by simple scaling. Assuming a double Fourier expansion of $U$ in space on a $[0;2 \pi)^2$ periodic domain, we can look at a single mode (single wavenumber) to understand the action of $\mathcal{L}$ in terms of its exponentials.
For a fixed time, let $U$ be of the form
 $U_{\vec{k}}(\vec{x})=e^{i \vec{k}\cdot \vec{x}}  \hat{U}_{\vec{k}},
 $ 
with $\vec{k}=(k_1,k_2)$, $\vec{x}=(x_1,x_2)=(x,y)$, $\hat{U}_{\vec{k}}$ independent of $\vec{x}$ and $i=\sqrt{-1}$.
Then 
 \begin{equation}
 \mathcal{L}U_{\vec{k}}=\left(\begin{array}{ccc}
  0 & f & -gik_1\\
 -f & 0 & -gik_2\\
 -\bar{\eta} i k_1& -\bar{\eta} i k_2 & 0  
\end{array}\right) \hat{U}_{\vec{k}},
 \end{equation}
  where the matrix above is the  matrix symbol of $\mathcal{L}$ (usually denoted as $\mathcal{L}(i\vec{k})$), 
  with 
  purely imaginary eigenvalues  \cite{majda2003introduction},
  \begin{equation}
\omega_f(\vec{k})=0, \quad  \omega_g(\vec{k})=\pm i\sqrt{f^2 + g \, \bar{\eta}\, \vec{k}\cdot\vec{k}}, 
\label{eq:swenormalmodes}
  \end{equation}
  where $\omega_f(\vec{k})$ is the steady geostrophic mode and $\omega_g$ defines the 2 inertia-gravity wave modes ($ \omega_g^{\,-}(\vec{k})$, $ \omega_g^{\,+}(\vec{k})$).
  The eigenvectors can be directly computed from $\mathcal{L}(i\vec{k})$, yielding a matrix of eigenvectors $Q$.
Writing the eigenvalues as a diagonal matrix
$\Lambda=[\omega_f(\vec{k}), \omega_g^{\,-}(\vec{k}), \omega_g^{\,+}(\vec{k})]$, and using $\mathcal{L}(i\vec{k})=Q\Lambda Q^{-1}$, the exponential of $\mathcal{L}$  can be directly calculated for the shallow water system through its symbol as
   \begin{equation}
   e^{\mathcal{L}(i\vec{k})}=Qe^{\Lambda} Q^{-1},
   \label{eq:swe_exp}
   \end{equation}
   where the $e^{\Lambda}$ is the diagonal matrix with entries given by the exponential of the respective eigenvalues.

For the studies conducted in the present work, we exploit features from double Fourier spectral spatial discretization.
This allows us to compute the numerical matrix exponential directly from equation \eqref{eq:swe_exp},
providing an exponential ($\varphi_0$) of the linear operator accurate to machine precision.
To evaluate $\varphi_n(\Delta t L)$ functions (see Eq.\,\eqref{eq:varphi_n}), we use that in spectral space
\begin{equation}
	\varphi_n(\Delta t\, \mathcal{L}(i\vec{k})) = Q \varphi_n( \Delta t\,\Lambda) Q^{-1},
\end{equation}
hence computing $\varphi_n$ element-wise for each diagonal element in $\Lambda$.

We would like to emphasize that computing the exponential directly as discussed above is only possible because we are exploiting the orthogonality of Fourier basis on the bi-periodic domain. 
For the SWE on the sphere, we'd like to mention that a similar feature was recently also discovered by using spherical harmonics to exponentially time integrate gravity modes.

In many problems such simplification may not always be possible, requiring different matrix exponentiation techniques. Even though many approaches to calculate exponentials exist (see \cite{hochbruck2010exponential}), two approaches are currently most commonly researched in this context: Krylov subspace solvers and rational approximations.
Krylov solvers, such as those presented in \cite{hochbruck1997krylov}, are used in \cite{Clancy2013} and \cite{GAUDREAULT2016827} for the matrix exponentiation of a dynamic linearization of the shallow water system.
Furthermore, \cite{schreiber2017beyond} adopts a rational approximation based on \cite{Haut2015} for the rotating SWE on the plane, which is also used for the sphere in \cite{Schreiber2018nla,schreiber_2019_exp_nonlinearswe_sphere} with a global spectral spherical harmonics representation. This rational approximation approach calculates the matrix exponentials with a very high degree of parallelism, so the additional computational costs of the calculating such exponential may be absorbed by extra compute nodes to still reduce the time-to-solution. 

In this study we will use the analytical linear operator exponential described in equation \eqref{eq:swe_exp}. Therefore, we will not address the discussion of different exponentiation techniques with respect to the semi-Lagrangian exponential method.

\subsection{Dispersion properties and C-grids}

The linear SWE on an f-plane define a hyperbolic system formed by inertia-gravity (Poincar\'e) and geostrophic (steady) waves, with the dispersion relations described in the previous section ($\omega_g$ and $\omega_f$, respectively). Numerical schemes should be able to represent well these two kinds of waves. We will adopt in this study spectral spatial discretizations of the linear operator (based on Fourier series), therefore errors in the evaluation of the linear operator are negligible (of machine precision) for each wavenumber. However, the temporal discretization may still be a source of error which can be directly investigated. 


Linear exponential integration schemes do not introduce any errors in time if the linear operator and its exponential is calculated analytically. 
However, state-of-the-art weather forecasting systems do not usually adopt exponential integration schemes, but mostly Runge-Kutta schemes \cite{Skamarock2012} when explicit, or Crank-Nicolson (CN) \cite{Hortal2002} when implicit (see a complete description in \cite{mengaldo2018current}). To ensure large time-steps, implicit schemes are preferred, but, in this case, the dispersion relations are usually not very accurately attained for the smaller wave-modes (faster gravity waves). Durran \cite{Durran2010} shows that the approximate dispersion ($\tilde{\omega}$) of the
CN scheme preserves the steady geostrophic modes (for $\tilde{\omega}_f(\vec{k})=\omega_f(\vec{k})=0$). However, the gravity waves will have dispersion of the form,
\begin{equation}
\tilde{\omega}_g(\vec{k})=\omega_g(\vec{k})+\frac{\Delta t^2}{12}(\omega_g(\vec{k}))^3 + \mathcal{O}(\Delta t ^5),
\end{equation}
which is purely imaginary (the amplitude of the mode is not altered by the scheme), but the phase speed is affected. The odd powers of $\omega_g$ indicate that the additional terms (error) will always produce a reduction of the $\tilde{\omega}_g$ frequency, and this reduction will be larger the larger the wavenumber norm ($\vec{k}\cdot \vec{k}$), since it depends on $\omega_g(\vec{k})$. Therefore,  the error in the Crank-Nicolson method slows down the faster (larger wavenumber) inertia-gravity waves, which will be slower when larger time-step sizes are used.

For finite difference schemes the spatial errors significantly influence the dispersion relations. \cite{Randall1994} analyzes the effect of different discretizations on the shallow water waves dispersions. To preserve an adequate representation of the inertia-gravity waves and reduce computational modes arising from spatial discretizations, staggered grids are preferred. These are usually called C-grids in the geoscientific modeling community, and have the depth variable centered in the cell and the velocities given at the edges of cell, normal to the edge \cite{Arakawa1977}.
%
%
Since many modern atmospheric models that use non-regular C- grids are using finite-differences/volumes with explicit time integration, we will also consider this approach as reference in our experiments further in the paper.

\section{Numerical experiments}
\label{sec:num}
We will consider the following set of schemes to be analyzed:

\begin{itemize}
\item RK-FDC: Runge-Kutta second order in time with second order in space energy conserving finite differences discretization on a staggered C-grid \cite{sadourny1975dynamics}.
\item SL-SI-SETTLS: Semi-Lagrangian, semi-implicit (Crank-Nicolson) scheme using spectral discretization adapted from \cite{Hortal2002} to the plane. 
\item SL-EXP-SETTLS: Exponential version of SL-SI-SETTLS (see Section \ref{sec:sl-exp-settls}).
\item ETD2RK: Original ETD2RK scheme  with spectral space discretization (see Section \ref{sec:expint}).
\item SL-ETD2RK: Semi-Lagrangian version of ETD2RK (see Section \ref{sec:sl-etdrk}).
\item REF: Reference solution. Runge-Kutta forth order in time with small time-step and high resolution Eulerian spectral discretization (pseudo-spectral for all nonlinear terms, such as advection).
\end{itemize}

The schemes are connected in the following way. RK-FDC is a reference explicit scheme well-established for the solution of the SWE of very low cost per time-step, but restricted to smaller time-steps (CFL condition). SL-SI-SETTLS is the state-of-the-art scheme used in many global atmospheric dynamical cores, which we aim to compare to our semi-Lagrangian exponential schemes (SL-EXP-SETTLS, SL-EXP-ETD2RK). ETD2RK is a well-established exponential integration technique, which we aim to compare to our semi-Lagrangian version, SL-ETD2RK, considering the different treatment of the nonlinear advection. 

\subsection{Definitions of domain and parameters}
The experiments will be executed on a scenario that mimics the Earth's dimensions, and we will follow the standard spherical test case parameters defined in \cite{Williamson1992}. The domain is set to be $[0,L_x]\times[0,L_y]=[0, 2 \pi a]\times[0, 2 \pi a]$, where $a=6371.22\,\mathrm{km}$  is the Earth radius, with bi-periodic boundary conditions. The gravity acceleration constant is set to $g=9.80616\, \mathrm{ms^{-2}}$ and the Coriolis frequency constant is $f=2\Omega$, with $\Omega=7.292\times 10^ {-5} \, \mathrm{rad\cdot s^{-1}} $. The mean depth is $\bar{\eta}=10 \, \mathrm{km} $ so that the gravity wave speed is $c = \sqrt{g\bar{\eta}} \approx 313 \,\mathrm{ms^{-1}}$.

The experiments will be performed with a horizontal discretization of 512 spectral modes in each dimension. This corresponds to 768 physical grid points to avoid aliasing effects, which would result in a grid cell with a length of approximately $52\, \mathrm{km}$ in each coordinate. The exception is the reference solution (REF), for which we will use 1024 spectral modes per coordinate. Such high horizontal resolution was chosen in order to reduce the errors relative to spatial discretizations and allow a clearer comparison of the different time stepping schemes. The time-step sizes will vary according to the analysis to be investigated.

We will present results of errors in two metrics: maximum absolute error (MaxError) and root mean square error (RMSError), always for fixed integration time (timestamp). 
In case of mismatching resolutions, where point-wise comparison is not well defined, bi-cubic spline interpolation is used on the highest resolution result to obtain information on the lowest resolution grid. This lack of matching happens as we are using a collocated grid (A-grid in geophysical notation) for the reference solution (REF), with physical representation of the quantities considered in the center of the cell.

\subsection{Kinetic energy spectra}

The analysis of the energy spectra is deeply related to the study of turbulence in fluid dynamics models, which is well investigated for the atmosphere (e.g.\,\cite{lindborg1999can, koshyk2001horizontal}).
Here, we do not intend to do turbulence analysis, but rather use spectrum analysis to compare small-scale wave interactions of the different schemes.
Therefore, we will assume a simplified kinetic energy spectrum analysis.

The one-dimensional Discrete Power Density Spectra (\cite{plunian2013shell}) is obtained from the two-dimensional kinetic energy spectrum using the Fourier transformed velocities $(\hat{u}(\vec{k}), \hat{v}(\vec{k}))$, as
\begin{equation}
E_n=\sum_{n\leq \|\vec{k} \| < n+1}E_{\vec{k}}, \quad \text{where} \quad E_{\vec{k}}=\frac{1}{2}\left(\hat{u}(\vec{k})\, \hat{u}^*(\vec{k})+\hat{v}(\vec{k})\,\hat{v}^*(\vec{k})\right),
\end{equation}
where $\vec{k}=(k_1,k_2)$ represents a horizontal mode, $^*$ represents the complex conjugate, $\|\vec{k}\|=\sqrt{k_1^2+k_2^2}$, and $E_n$ represents the spectrum density with respect to the wavenumber $n$ and wavelength $L/n$, with $L$ the size of the domain. This closely follows what is usually done in spherical atmospheric models (e.g. \cite{koshyk2001horizontal}).

\subsection{Unstable jet test case}

On the sphere, a well-known test case is defined by the Galewsky et al \cite{Galewsky2004} initial conditions. These initial conditions define a geostrophically balanced mid-latitude zonal jet. A small perturbation in the height field is added in order to generate fast gravity waves that eventually destabilize the jet and form well-defined vortices after a few days.

On the bi-periodic plane, no such test case exists, so we propose something similar in the following way.
We define two jets by the $u$ and $v$ velocities as, 
\begin{equation}
u(x,y)=u_0 \left(\sin(2 \pi y/L_y)\right)^{81},\quad\quad v(x,y)=0,
\end{equation}
where $u_0=50 \mathrm{ms^{-1}}$ is the maximum speed and the power of $81$ was chosen so that the jet is confined in a small region. To ensure that the depth field is in balance with the velocity field, that is, that the initial conditions are analytically in a steady state, we define the depth perturbation as
\begin{equation}
\eta(x,y) = -\frac{f}{g} \int_{0}^y u(x,s) ds.
\end{equation}
The integral is solved numerically through repeated piece-wise Gaussian integrals ensuring that the integral is calculated within desired tolerance for double precision. 

Small Gaussian perturbations ($\eta_p$) are added to $\eta$ to trigger the barotropic instability, 
\begin{equation}
\eta_p(x,y) = 0.01 \bar{\eta} \left[\exp\{-k d_1(x,y))\} + \exp\{-k d_2(x,y)\}\right],
\end{equation}
where $k=1000$, and $
d_i(x,y)=\frac{(x-x_i)^2}{L_x^2}+\frac{(y-y_i)^2}{L_y^2}
$, $i=\{1,2\}$, 
are the square Euclidean distances of $(x,y)$ to the points $p_1=(x_1,y_1)=(0.85L_x,0.75L_y)$, $p_2=(x_2,y_2)=(0.15 L_x,0.25 L_y)$, respectively. 

\begin{figure}[!ht]
	\centering
	(a)\includegraphics[scale=0.28]{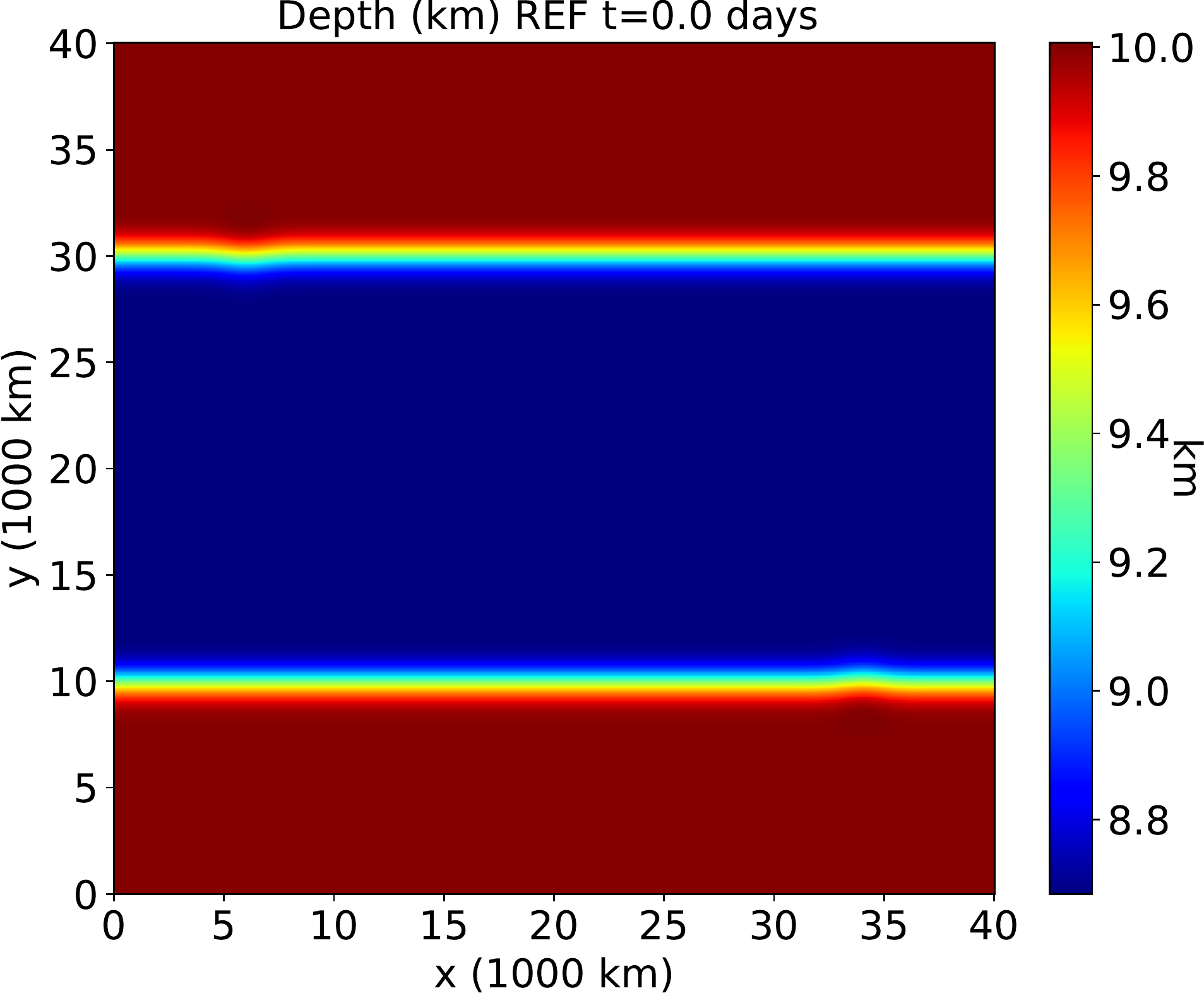}
	\hspace{0.5cm}
	(b)\includegraphics[scale=0.28]{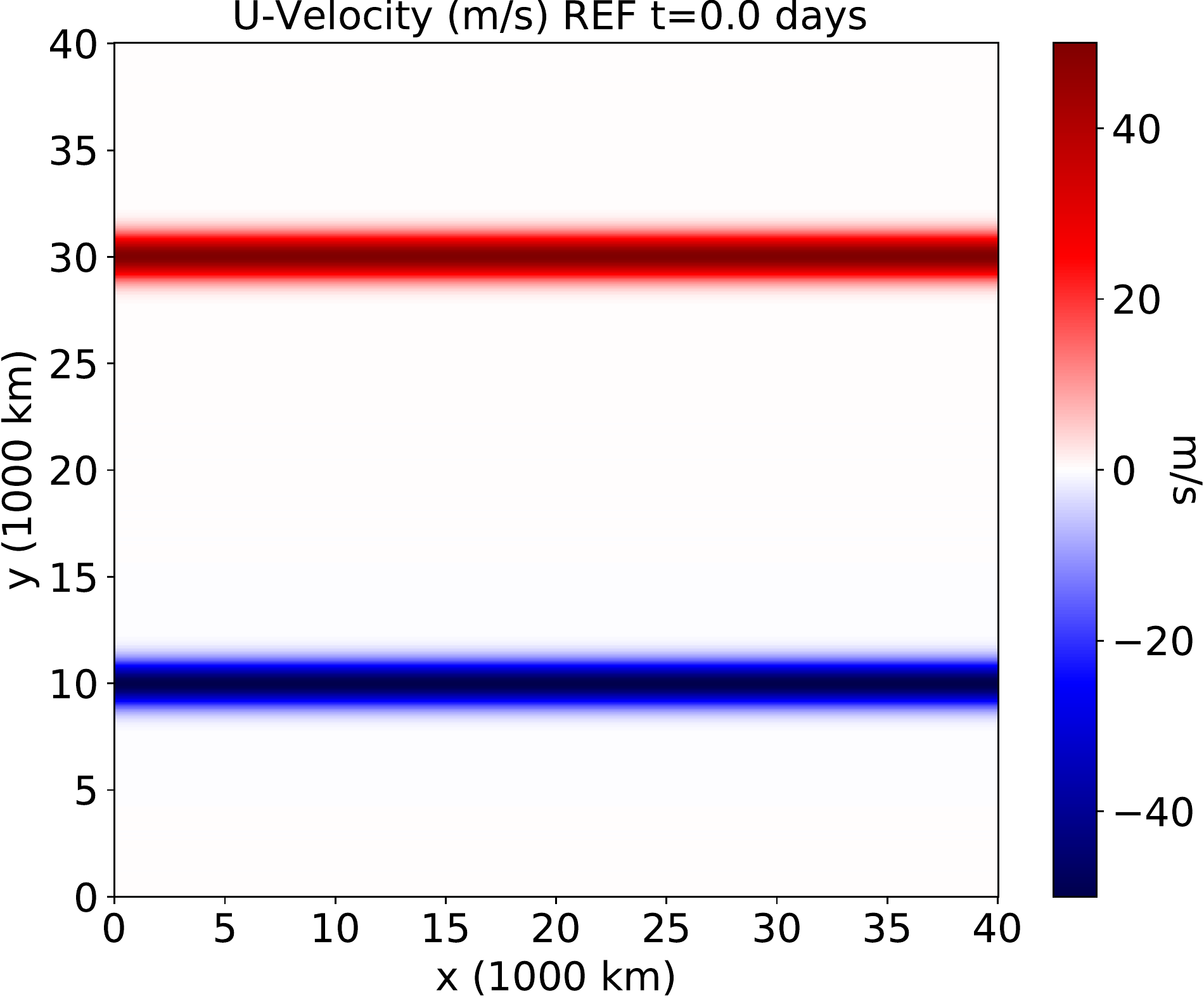}
	\caption{Initial conditions for the unstable jet test case. (a) Total depth ($\eta+\bar{\eta}$) and (b) zonal velocity ($u$).}
	\label{fig:unstablejet_ic}
\end{figure}

\begin{figure}[h!]
\centering
(a)\includegraphics[scale=0.28]{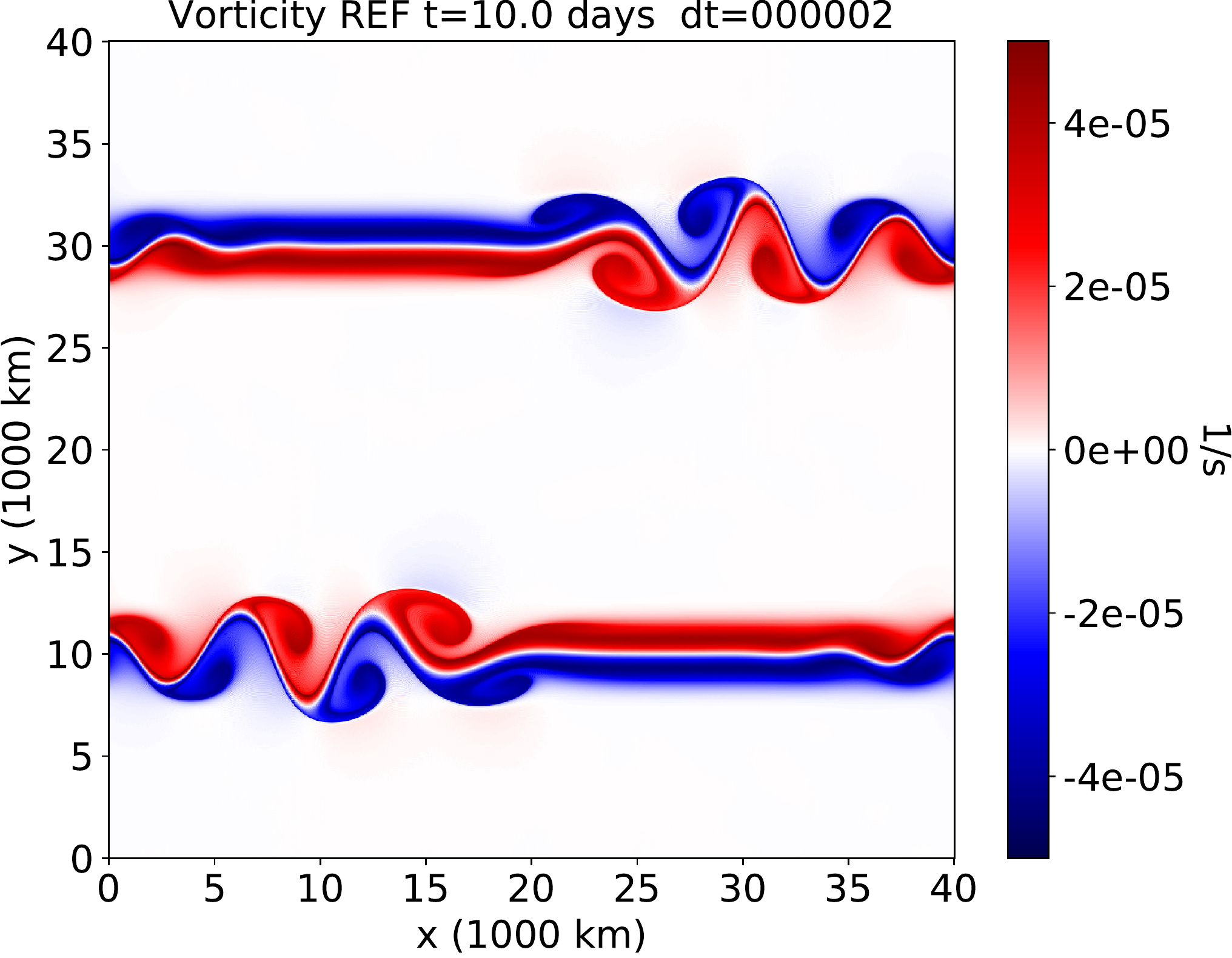}
(b)\includegraphics[scale=0.28]{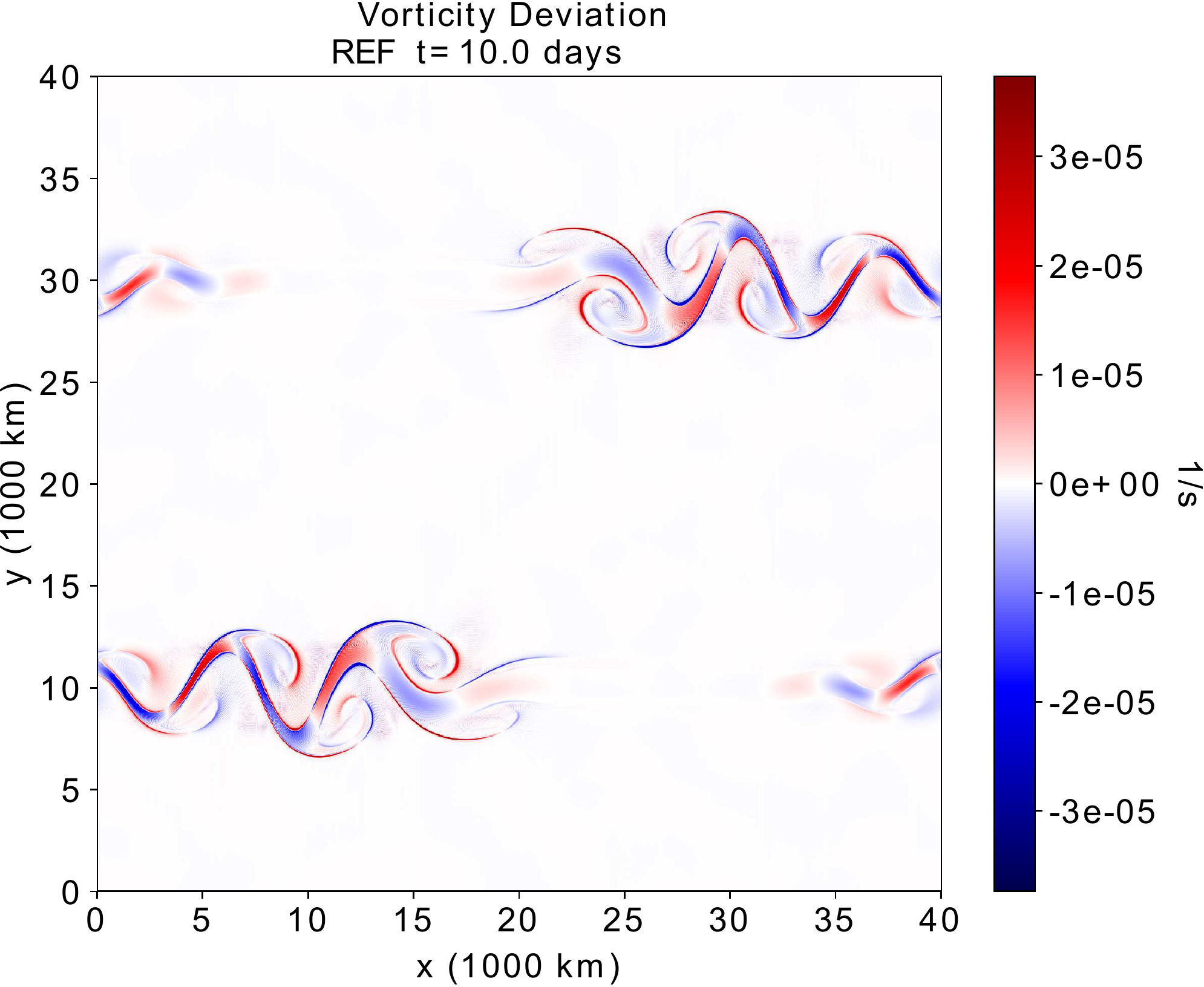}\\
\caption{Reference solution for vorticity at 10 days. (a) Full nonlinear SWE. (b) Difference from full SWE to the one with neglected nonlinear divergence term ($\tilde{\mathcal{N}}=0$).}
\label{fig:unstable_ref_10}
\end{figure}

Initial conditions are presented in Figure \ref{fig:unstablejet_ic}. Note that the zonal jets move towards different directions (left-right), in order to ensure periodicity of all initial fields.
We present in Figure \ref{fig:unstable_ref_10} results from the high resolution reference scheme (REF) with a small time-step size of 2 seconds.
The initial Gaussian perturbations trigger the generation of fast-moving inertia-gravity waves that dominate the initial period of time integration.
The waves start interacting with each other through the nonlinear effects and eventually disturb the jets to form well-defined vortices at day 10, shown in Figure \ref{fig:unstable_ref_10}(a) with the vorticity of the flow.

We will also use this test case neglecting the nonlinear divergence of the SWE ($\tilde{\mathcal{N}}$ from equation \eqref{eq:swe}). The SWE flow is still nonlinear, due to the nonlinear advection term. In fact, the solution of the unstable jet initial condition  neglecting the nonlinear divergence is very similar to the solution considering this term (see Figure \ref{fig:unstable_ref_10}b). Even though this term might not visually influence significantly the solutions after 10 days, it plays an important role in energy cascade and nonlinear interaction of waves. Also, it will influence the numerical properties of the scheme, as we will see further on in the next section. 

\begin{figure}[h!]
\centering
(a)\includegraphics[scale=0.28]{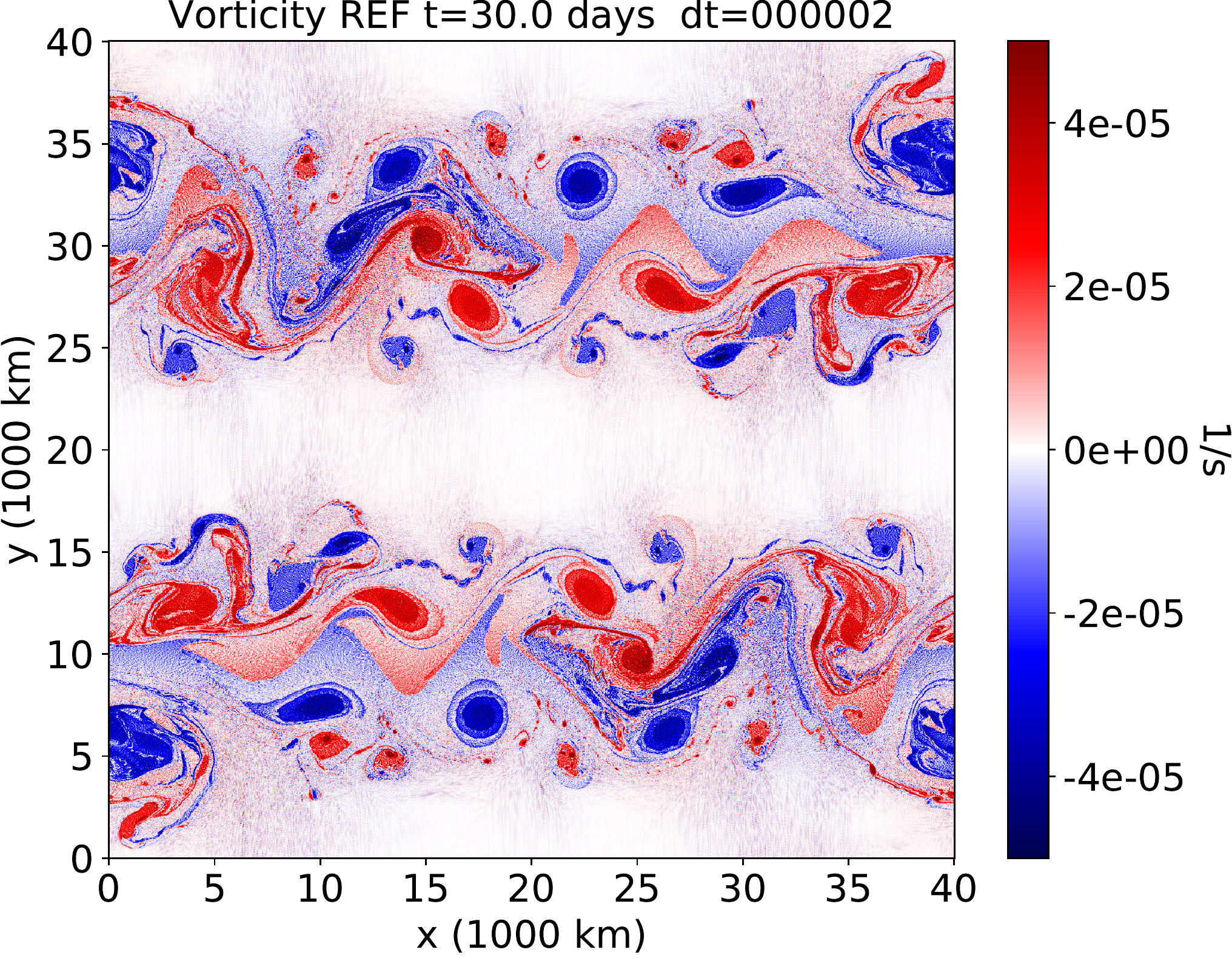}
(b)\includegraphics[scale=0.28]{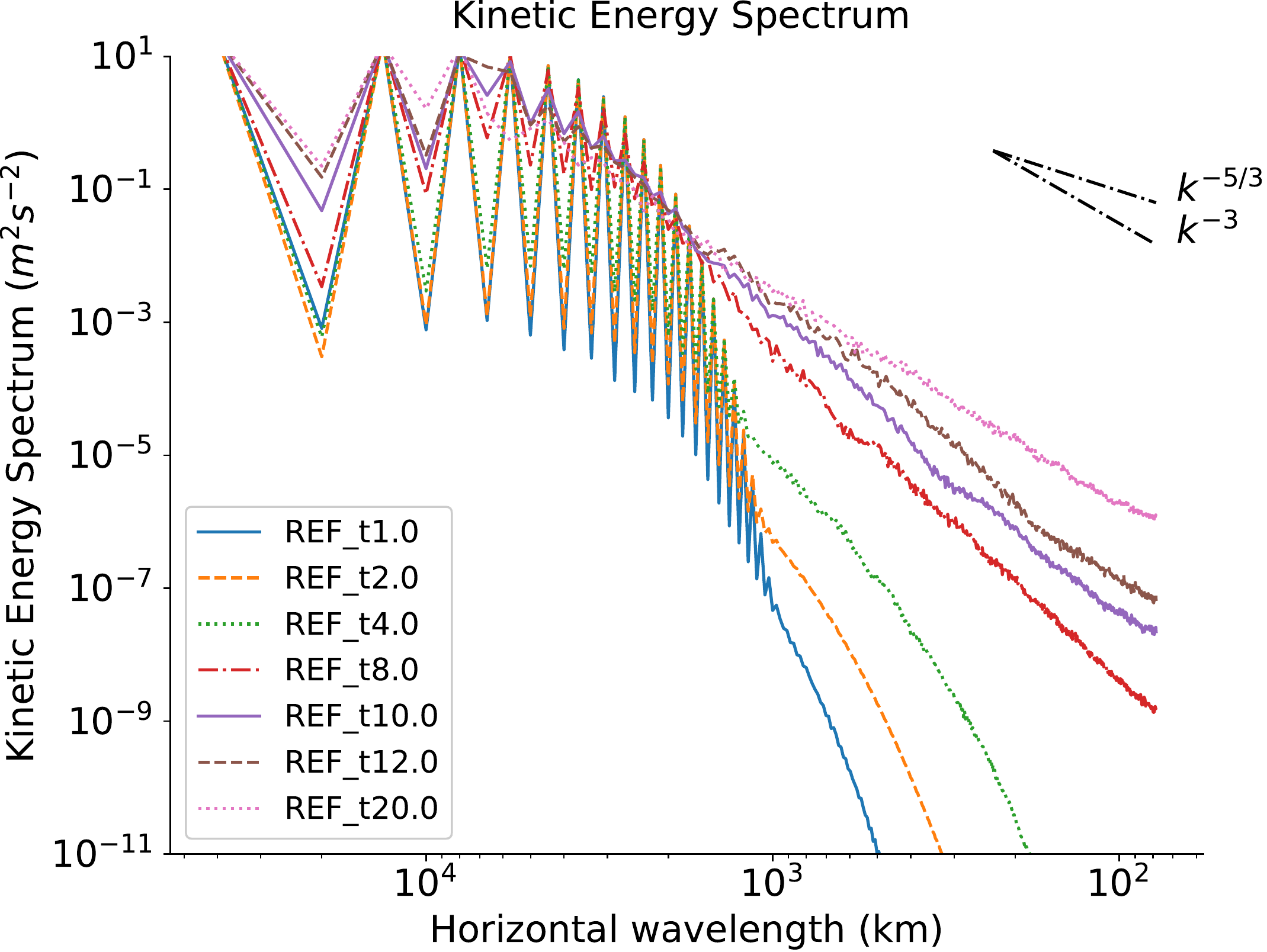}\\
\caption{(a) Reference solution for vorticity at 30 days using the full nonlinear SWE. (b) Kinetic energy spectrum for reference solution using the full nonlinear SWE for different integration times (from 1 day to 20 days).}
\label{fig:unstable_ref_30}
\end{figure}



After longer periods of time, the flow develops into a fully turbulent regime, as may be seem in Figure \ref{fig:unstable_ref_30}a (the flow considering $\tilde{\mathcal{N}}=0$ is very similar to the full SWE). From a spectral point of view, energy moves towards smaller wavelengths as time evolves, as may be seen in Figure \ref{fig:unstable_ref_30}b. The initial kinetic energy spectrum is basically defined by the spectrum of powers of trigonometric functions  (in this case $\sin^{81}(2 \pi y/L_y)$ ). As the power chosen (81) is odd, the spectrum will be zero for all even wavenumbers.
That is why we see a zig-zag pattern in the early stages of integration in the kinetic energy spectrum for $t>0$.
Energy builds up in even wavenumbers due to nonlinear interactions. Note also that the spectra converges towards the well known $-5/3$ power law of 2D kinetic energy turbulence \cite{lindborg1999can}. Reproducing this kind of spectra in small wavelengths stably is usually a major challenge for numerical schemes.

\subsection{Analysis of the Shallow Water Equations without nonlinear divergence}

Considering $\tilde{\mathcal{N}}=0$ simplifies the semi-Lagrangian exponential schemes. In fact, in this case, SL-EXP-SETTLS and SL-ETD2RK are equivalent, since the only non-linearity left (advection) is treated within the semi-Lagrangian approach. SL-SI-SETTLS  also greatly simplifies for similar reasons. RK-FDC, ETD2RK and REF still have to deal with the nonlinear advection as a nonlinear term. The finite differences scheme RK-FDC is built about the vector invariant form of the equations, where nonlinear advection is not explicit, therefore it is not clear how to remove the nonlinear divergence and we do not present results of this scheme in this case.

\begin{figure}[!ht]
	\centering
	(a)\includegraphics[scale=0.26]{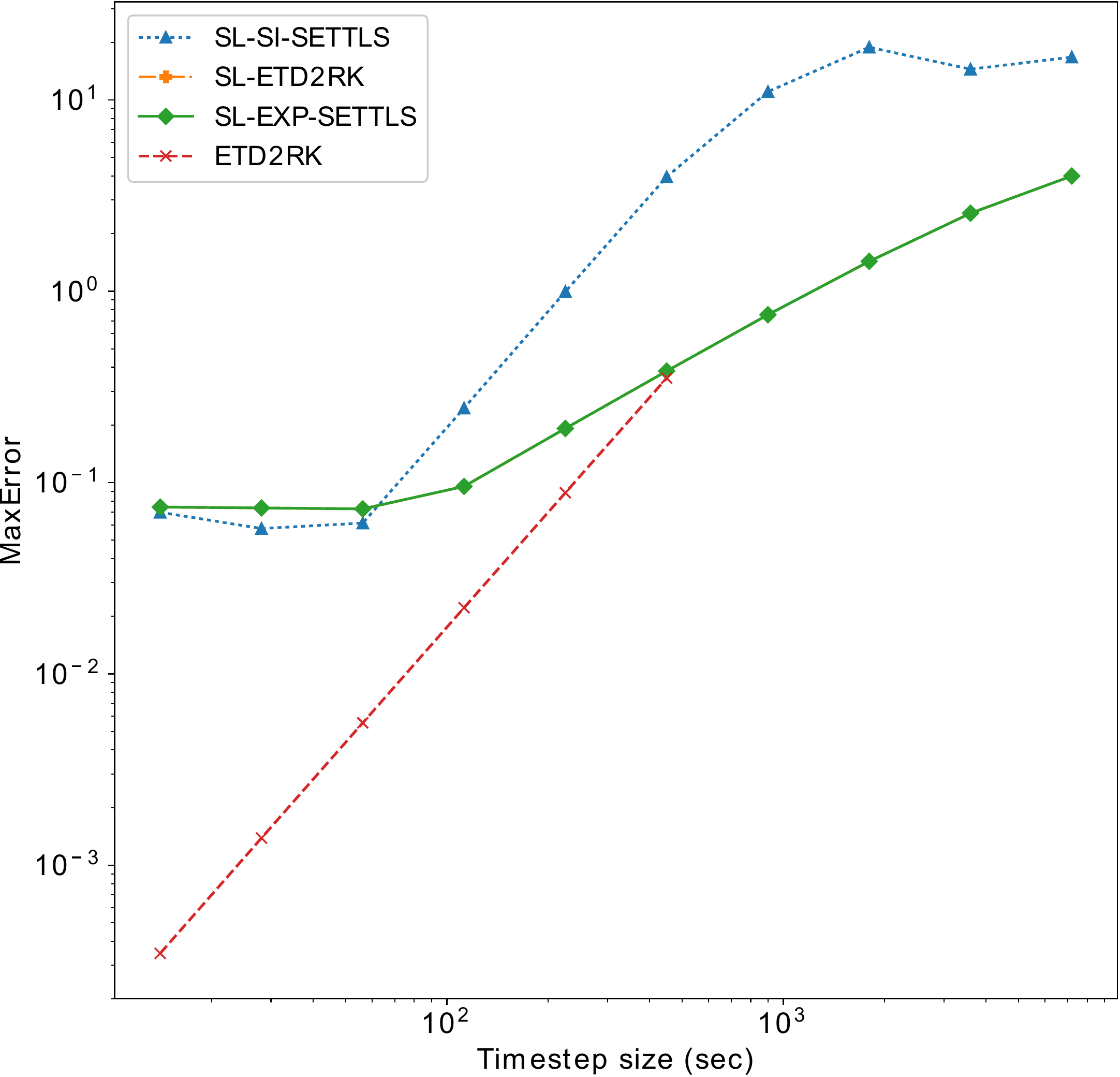}
	(b) \includegraphics[scale=0.26]{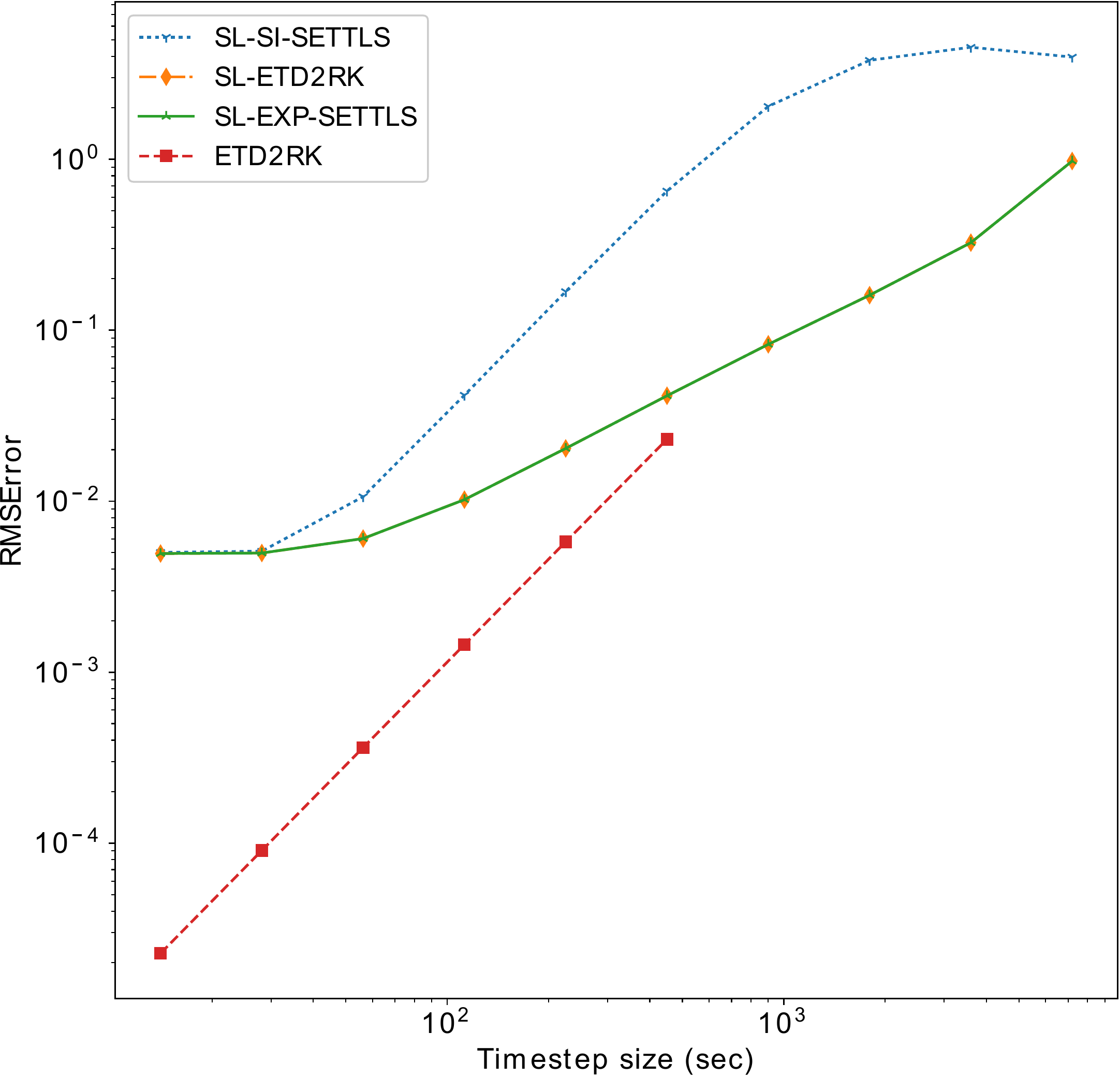}
	\caption{Errors at day 1 of integration of the
		unstable jet test case without nonlinear divergence. (a) Maximum absolute error and (b) RMS error for day 1 of integration with respect to the reference solution (REF) for different time-step sizes and schemes. All schemes were tested for all time-step sizes indicated. If a scheme does not shows a value for a large time-step size it indicates that it became unstable for this test. The SWE without nonlinear divergence were adopted in this test, therefore SL-EXP-SETTLS and SL-ETD2RK are identical.}
	\label{fig:swe_conv}
\end{figure}

The initial period is dominated by linear gravity waves, so that is where we expect to see benefits of the exponential integration scheme with respect to the semi-implicit scheme.
We show in Figure \ref{fig:swe_conv} the errors at day 1 of integration for the unstable jet test case without nonlinear divergence.
It should be noted that for small time-step sizes
the dominating error in the semi-Lagrangian schemes becomes the spatial interpolation errors, not the temporal.
These errors for small time-step sizes may be reduced by considering the resolution proportional to the time-step size ($\Delta x \propto \Delta t$) or increasing the accuracy order of the semi-Lagrangian scheme.
For smaller time-step sizes, the ETD2RK is the most accurate one, since ETD2RK has all spatial operators treated spectrally, hence no interpolation errors.

However, the semi-Lagrangian schemes are stable throughout all time-step sizes tested, whereas the ETD2RK scheme is limited by advection CFL time-step size. In general, the semi-Lagrangian exponential schemes are more accurate than the semi-implicit scheme (SL-SI-SETTLS), due to the accurate treatment of the linear waves. Concluding, the semi-Lagrangian exponential schemes provide a more accurate way, compared to SL-SI-SETLLS, to extend the time-step size allowed by the traditional exponential scheme (ETD2RK).

\begin{figure}[!ht]
	\centering
	\includegraphics[scale=0.30]{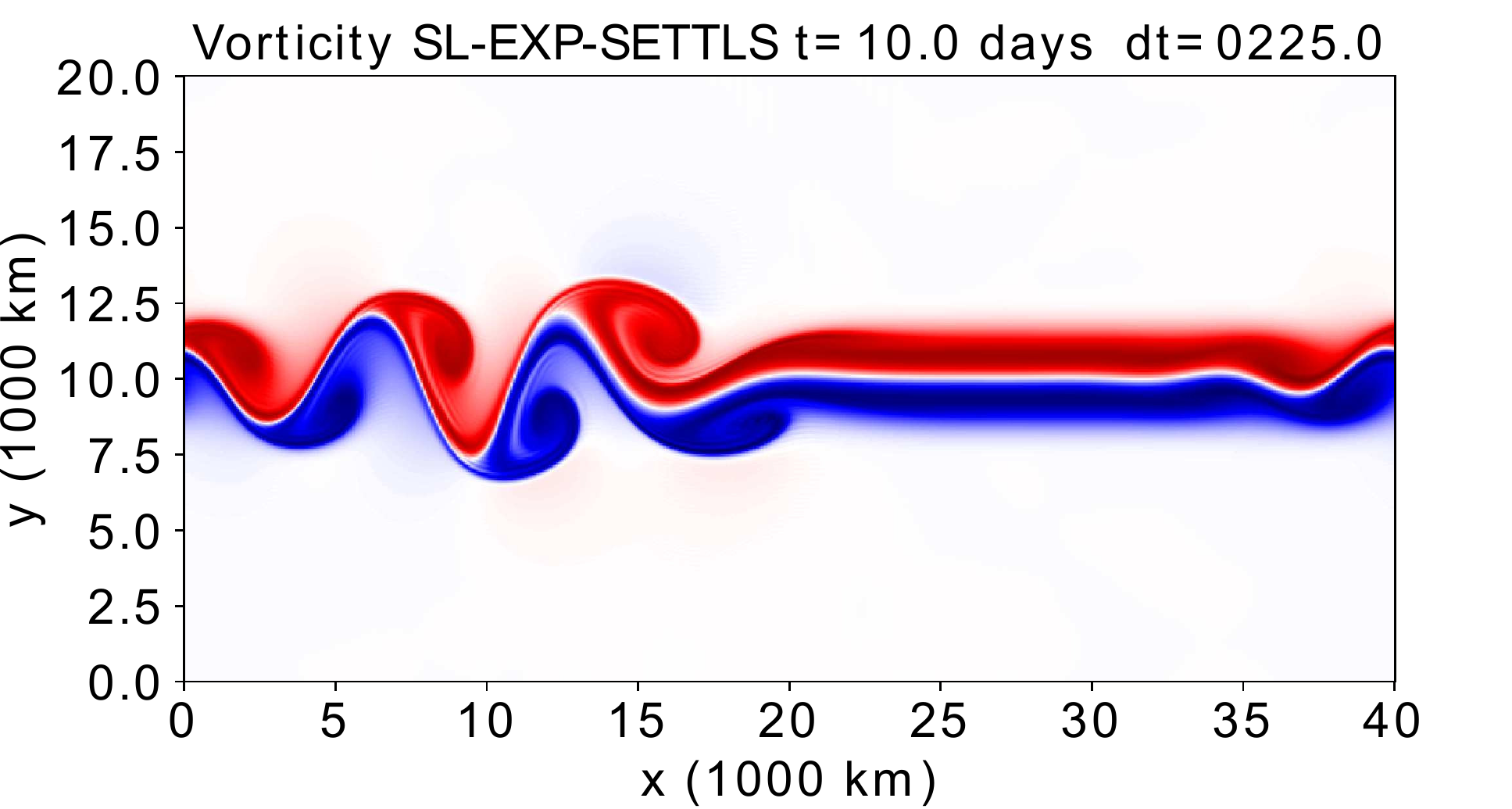}
	\zoomedareaimage{figures/{unstablejetlinearscript_sl-rexi_g9.80616_h10000_f0.00014584_u0.0_U0_tsm_l_rexi_na_sl_nd_settls_tso2_tsob2_C0225.0_REXIDIR_M0512_diag_vort_t00000864000.00000000halfcut}.pdf}
	\includegraphics[scale=0.30]{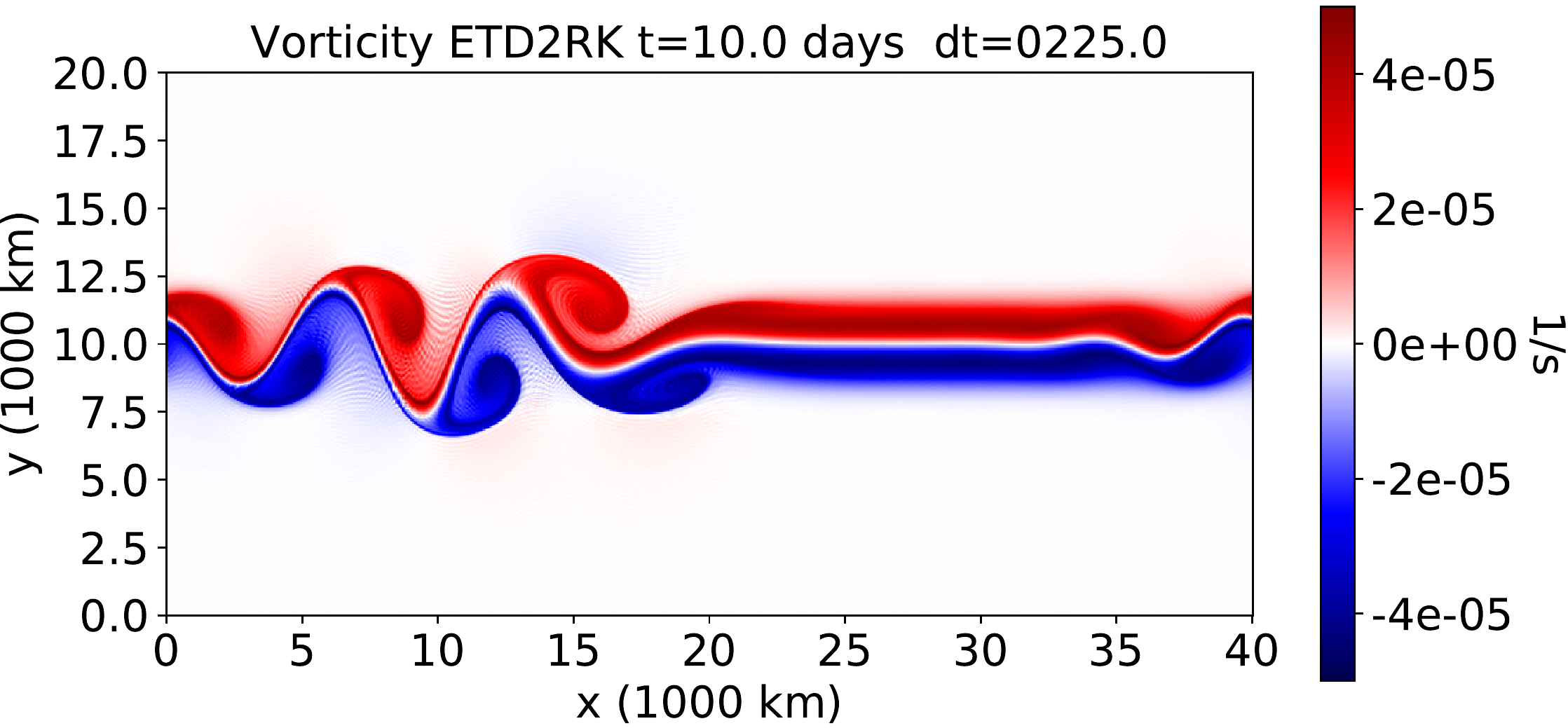}
	\zoomedareaimagex{figures/{unstablejetlinearscript_sl-rexi_g9.80616_h10000_f0.00014584_u0.0_U0_tsm_l_rexi_n_etdrk_tso2_tsob2_C0225.0_REXIDIR_M0512_diag_vort_t00000864000.00000000half}.pdf}
	\\
	\begin{flushleft}
		\vspace{-0.7cm}
		(a) \hspace{6cm} (b)
	\end{flushleft}
	\caption{Numerical solution of the SWE without nonlinear divergence for the unstable jet test case at the time of 10 days for the vorticity field using a time-step size of 225 seconds. The box within each figure shows an amplification of the main vortex formation.	(a) SL-SI-SETTLS/SL-EXP-SETTLS/SL-ETD2RK (identically looking, only plotted one for sake of brevity), (b) ETD2RK. We can observe more small scale features for the ETD2RK method around the vortex borders.}
	\label{fig:swe_vorticity225sec}
\end{figure}


Due to the dynamically unstable (chaotic) nature of the test case, quantitative analysis of errors in longer periods of time is not usually indicated. However, it is interesting to see qualitatively how the schemes behave once the vortices have developed. We show in Figure \ref{fig:swe_vorticity225sec} the vorticity at day 10 for the schemes investigated. All schemes seem to be able to represent well the vortex formation, but we notice that the ETD2RK has more oscillations at or around the vortices, whereas the semi-Lagrangian schemes show smoother vortices.

\subsection{Analysis of the Full Shallow Water Equations}

\begin{figure}[!ht]
	\centering
	(a)\includegraphics[scale=0.27]{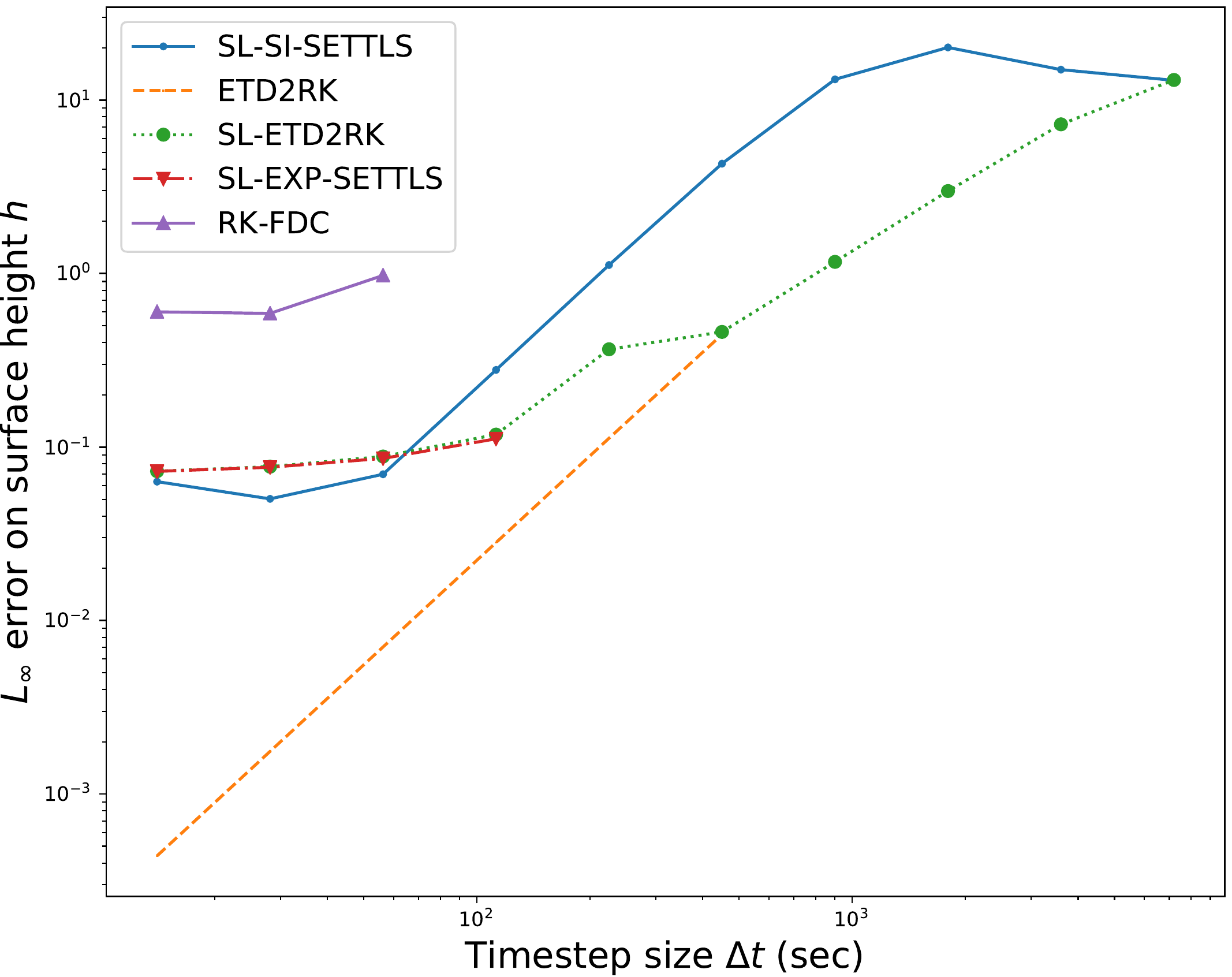}
	(b) \includegraphics[scale=0.27]{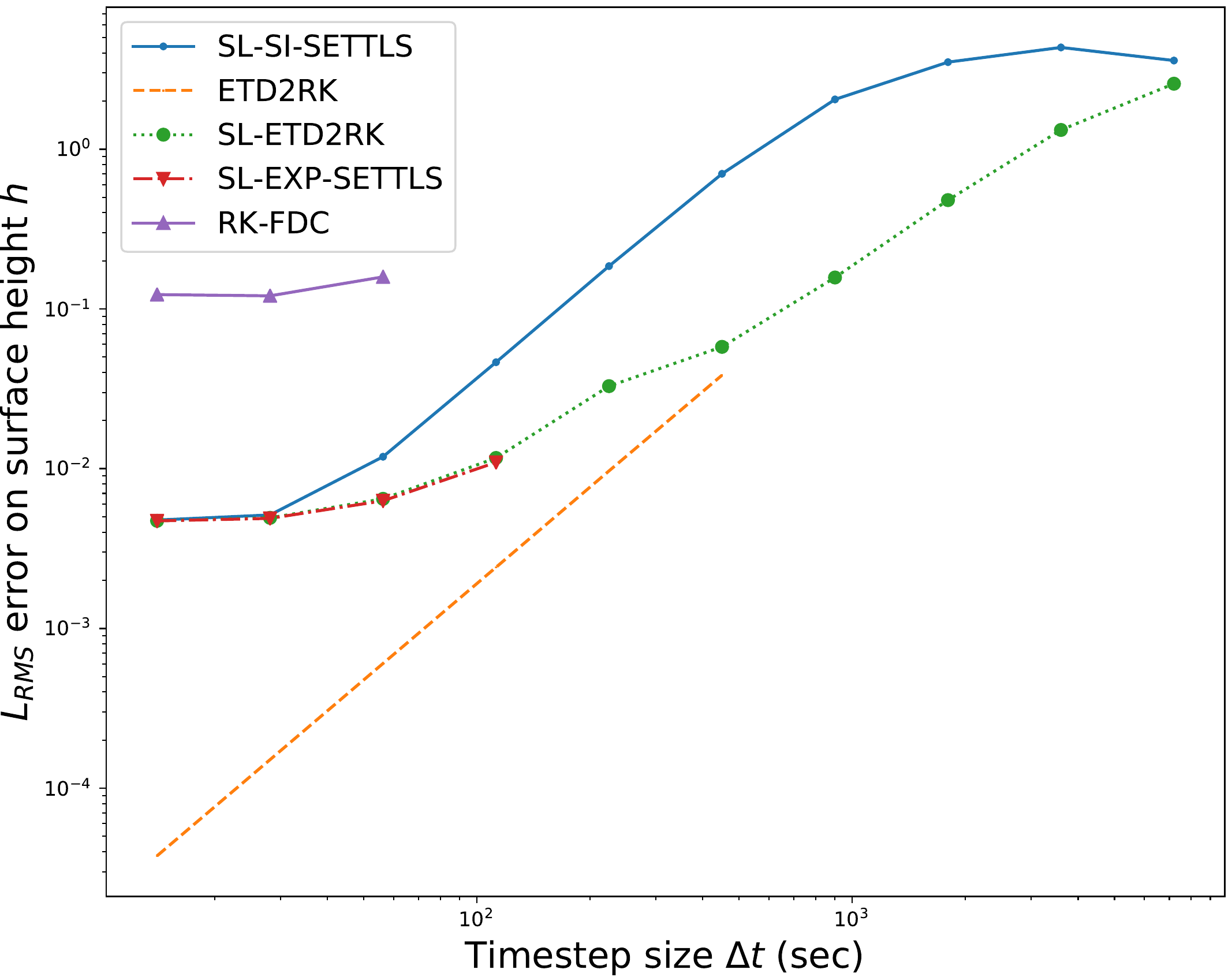}
	\caption{Errors on surface height at day 1 of integration of the unstable jet test case for the full nonlinear SWE.
		(a) Maximum absolute error and (b) RMS error.
		Same as Figure \ref{fig:swe_conv} but now considering the nonlinear divergence.}
	\label{fig:full_swe_conv}
\end{figure}

In this section, we will analyze the schemes with respect to the full SWE, including the nonlinear divergence. In this case, the RK-FDC schemes will also be included in the analysis. Also, the different semi-Lagrangian exponential schemes (SL-ETD2RK and SL-EXP-SETTLS) now differ from each other.

%
%
We start by studying the results of all time integration methods at day 1 of the full SWE for the unstable jet test case.
Convergence plots for varying time-step sizes are presented in Figure \ref{fig:full_swe_conv}.
As in the previous test, due to the limitation imposed by the spatial interpolation used in the semi-Lagrangian schemes, the ETD2RK scheme provides more accurate results for smaller time-step sizes.
The ETD2RK scheme is again restricted in large time-step sizes by CFL condition for advection.
The RK-FDC scheme is limited in both time and space: the finite differences scheme limits the accuracy, and the gravity wave speed CFL limits the time-step size.
With the inclusion of the nonlinear divergence, the SL-EXP-SETTLS scheme turns out to be unstable when used with large time-steps. Compared to the SL-SI-SETTLS scheme, the SL-EXP-SETTLS preserves better
the high wavenumber gravity waves, which interact with each other in the nonlinear divergence and become numerically unstable. Differently, the SL-ETD2RK scheme is stable with large time-steps, and is more accurate than the SL-SI-SETTLS scheme, due to the more accurate treatment of the linear waves. The theoretical stability analysis of the semi-Lagrangian schemes is still a matter to be investigated and is here considered only in an empirical sense. However, we point out an important difference between them: SL-EXP-SETTLS is a multistep scheme (requires an extrapolation from a previous time-step), whereas the SL-ETD2RK is a single step method (apart from the extrapolation used in the back trajectory calculation).
We also notice that SL-ETD2RK seems to be a viable extension of the ETD2RK scheme to larger time-steps, being more accurate than the SL-SI-SETTLS.

\begin{figure}[!ht]
	\centering
	\includegraphics[scale=0.3]{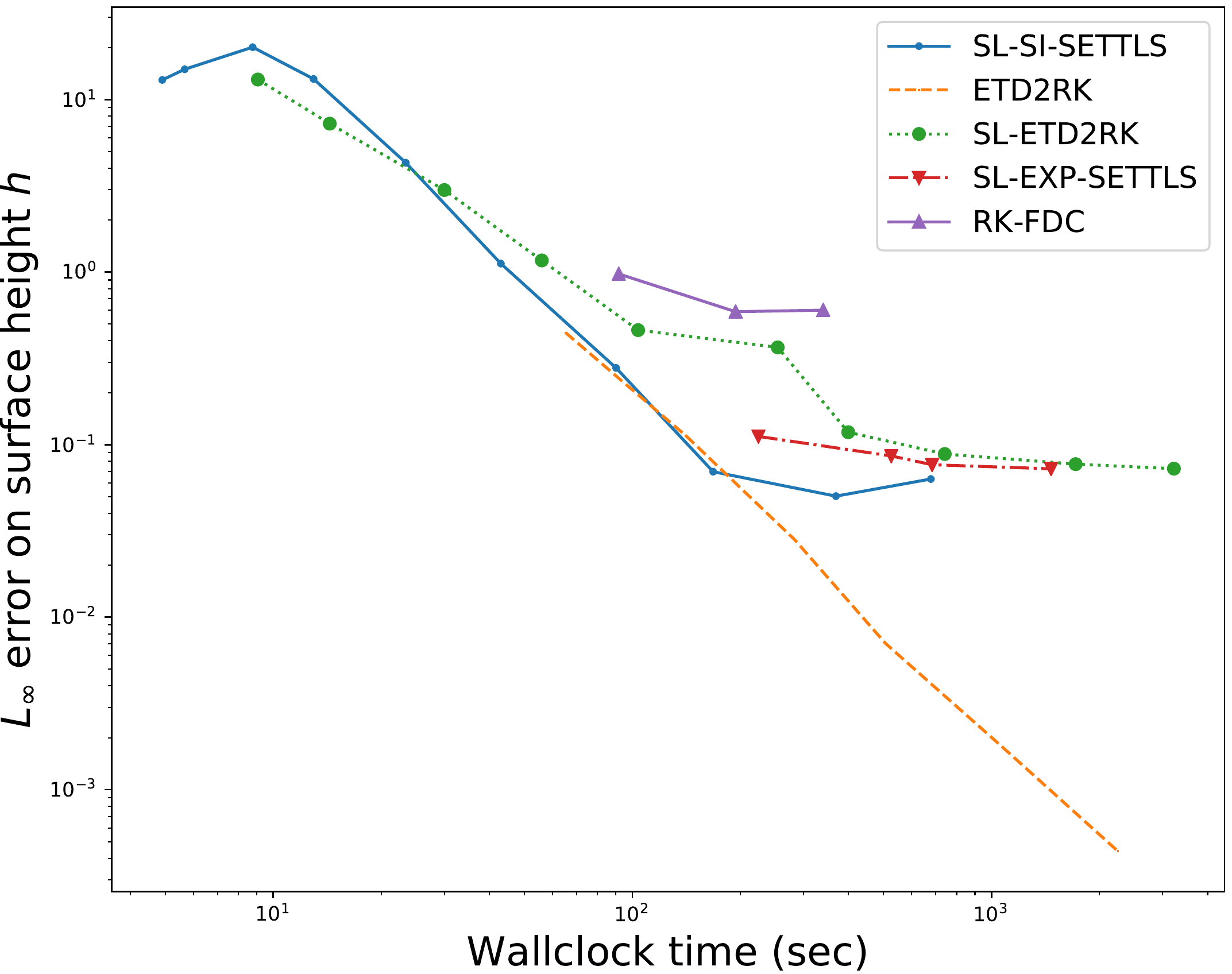}
	\caption{Errors vs.\,wallclock time at day 1 for different time integration methods. Assuming the request to stay below 100s of wallclock time, we can observe that the SL-SI-SETTLS and SL-EXP-ETDRK methods provide the best results.}
	\label{fig:full_swe_error_vs_wallclock_time}
\end{figure}

So far, we only compared the error with the time-step size of individual time integration methods.
However, the time-step size does not directly relate to the total computational requirements and therefore also not to the wallclock time.
Although it is challenging to run representative wallclock times studies which also relate to a fully-developed dynamical core running on a super computer, we provide such wallclock time studies in Figure \ref{fig:full_swe_error_vs_wallclock_time}. Small wallclock times indicate larger time-step sizes.
Here, we can observe that the exponential semi-Lagrangian scheme SL-ETD2RK has competitive wall-clock times compared to the state-of-the-art SL-SI-SETTLS method, particularly for small wall-clock times. There are, nevertheless, additional computing requirements of the exponentials, but we would like to point out that these are the first types of SL-ETDnRK methods and we expect improved results with further computational, mathematical and modeling optimizations.
Again, we also observe a limitation of the accuracy for all semi-Lagrangian methods for small time-step sizes, due to the 2nd order accuracy in space. For high accuracy regimes, with small time-step sizes, the ETD2RK scheme provided the best results.

%
%
%
%

\begin{figure}[!ht]
	\centering
	\includegraphics[scale=0.3]{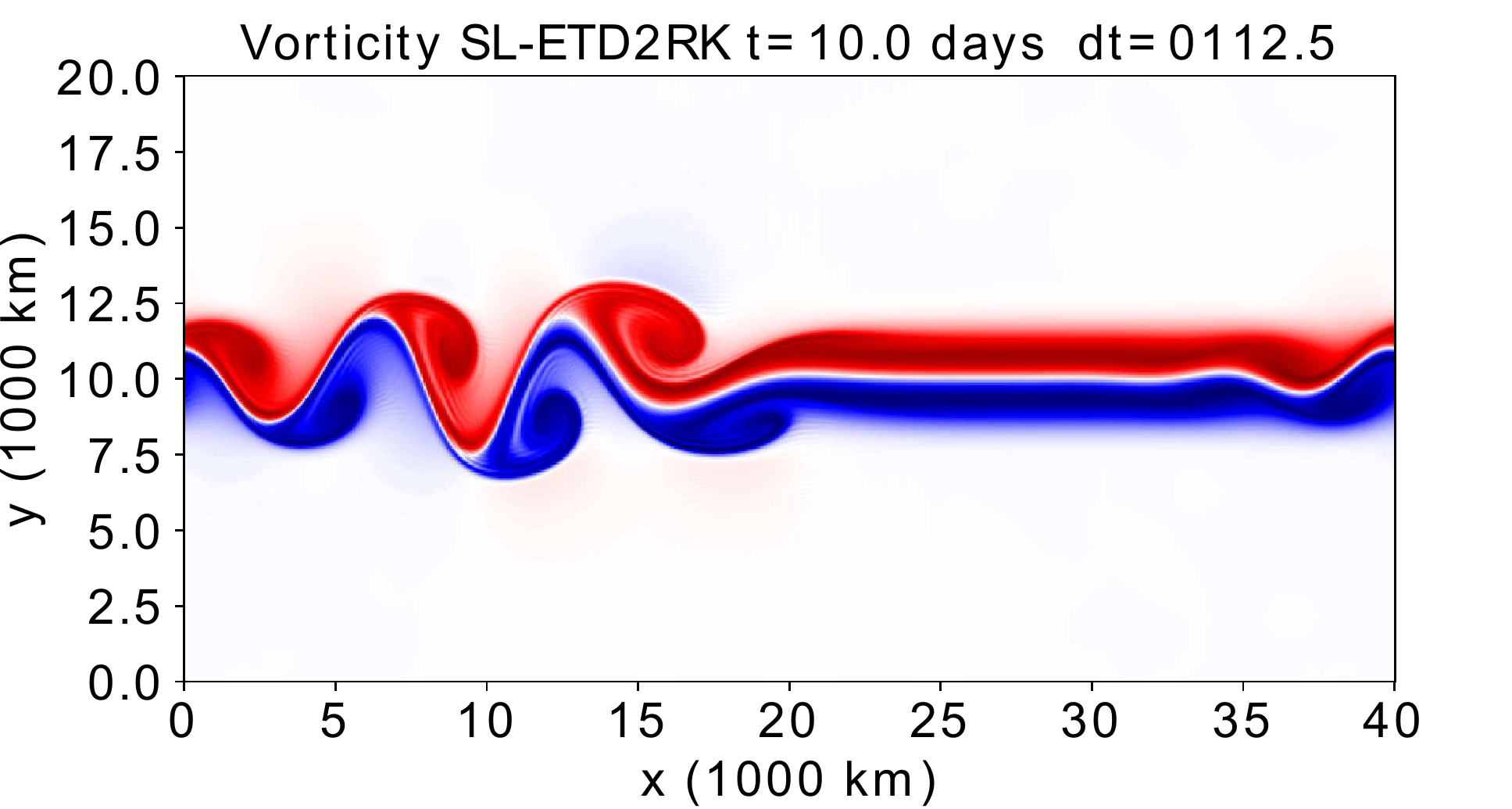}
	\zoomedareaimage{figures/{unstablejetscript_sl-rexi_g9.80616_h10000_f0.00014584_u0.0_U0_tsm_l_rexi_na_sl_nd_etdrk_tso2_tsob2_C0112.5_REXIDIR_M0512_diag_vort_t00000864000.00000000halfcut}.pdf}	
	\includegraphics[scale=0.3]{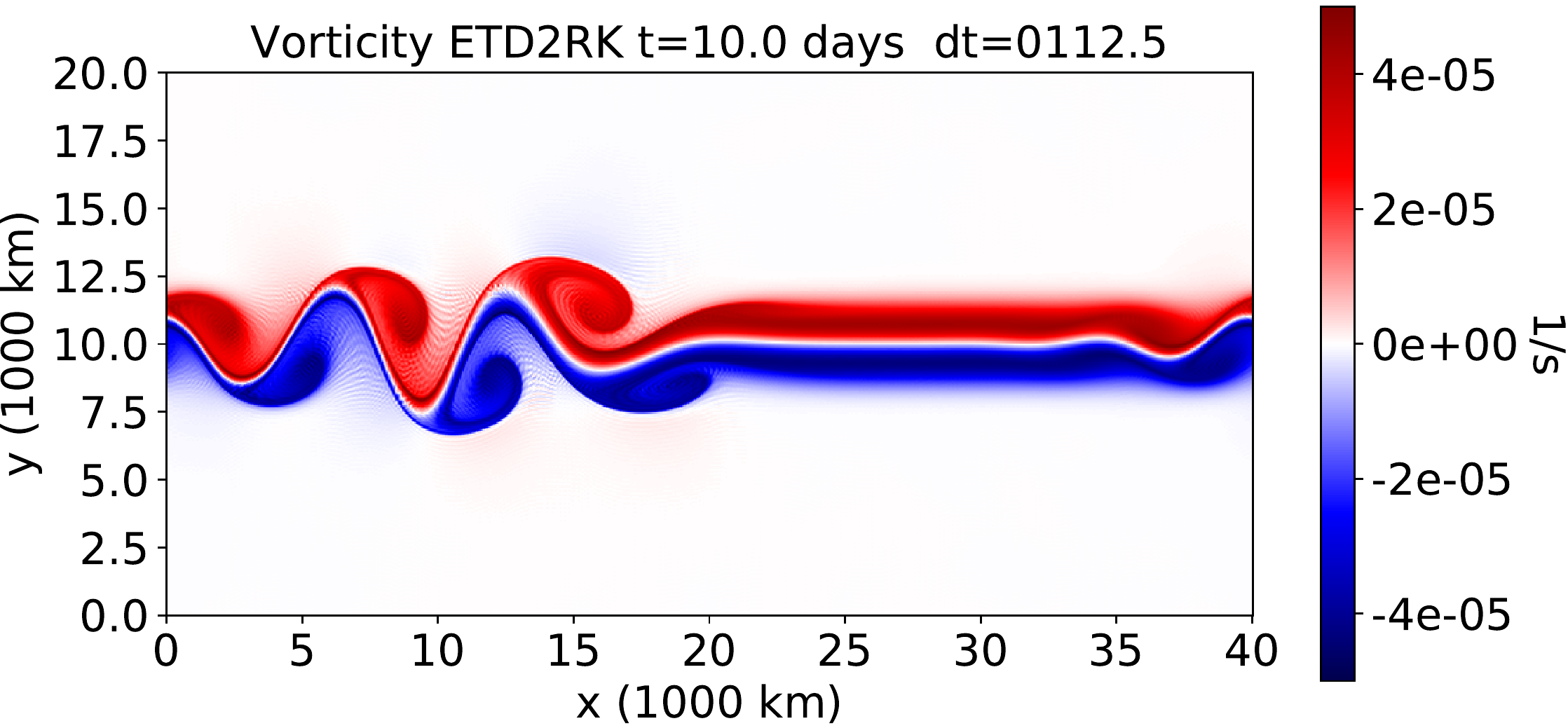}
	\zoomedareaimagex{figures/{unstablejetscript_sl-rexi_g9.80616_h10000_f0.00014584_u0.0_U0_tsm_l_rexi_n_etdrk_tso2_tsob2_C0112.5_REXIDIR_M0512_diag_vort_t00000864000.00000000half}.pdf}	\\
	\begin{flushleft}
		\vspace{-0.7cm}
		(a) \hspace{6cm} (b) 
	\end{flushleft}
	\caption{Vorticity field of the full nonlinear SWE for the unstable jet test case at day 10 using a time-step size of 112.5 seconds.
		Images refer to (a) SL-ETD2RK and (b) ETD2RK schemes.
		The plot for SL-SI-SETTLS is very similar to SL-ETD2RK and therefore not shown for sake of brevity.}
	\label{fig:fullswe_vorticity112sec}
\end{figure}

Now, we study results for different time integration methods at day 10 of the benchmark.
The vorticity fields for two different schemes (SL-ETD2RK, ETD2RK) are depicted in Figure \ref{fig:fullswe_vorticity112sec}, where, since SL-SI-SETTLS provides results similar to SL-ETD2RK, this scheme was not depicted.
They are again qualitatively very similar, although the ETD2RK shows more high wavenumber oscillations around the vortices.

\begin{figure}[!ht]
	\centering
	\includegraphics[scale=0.3]{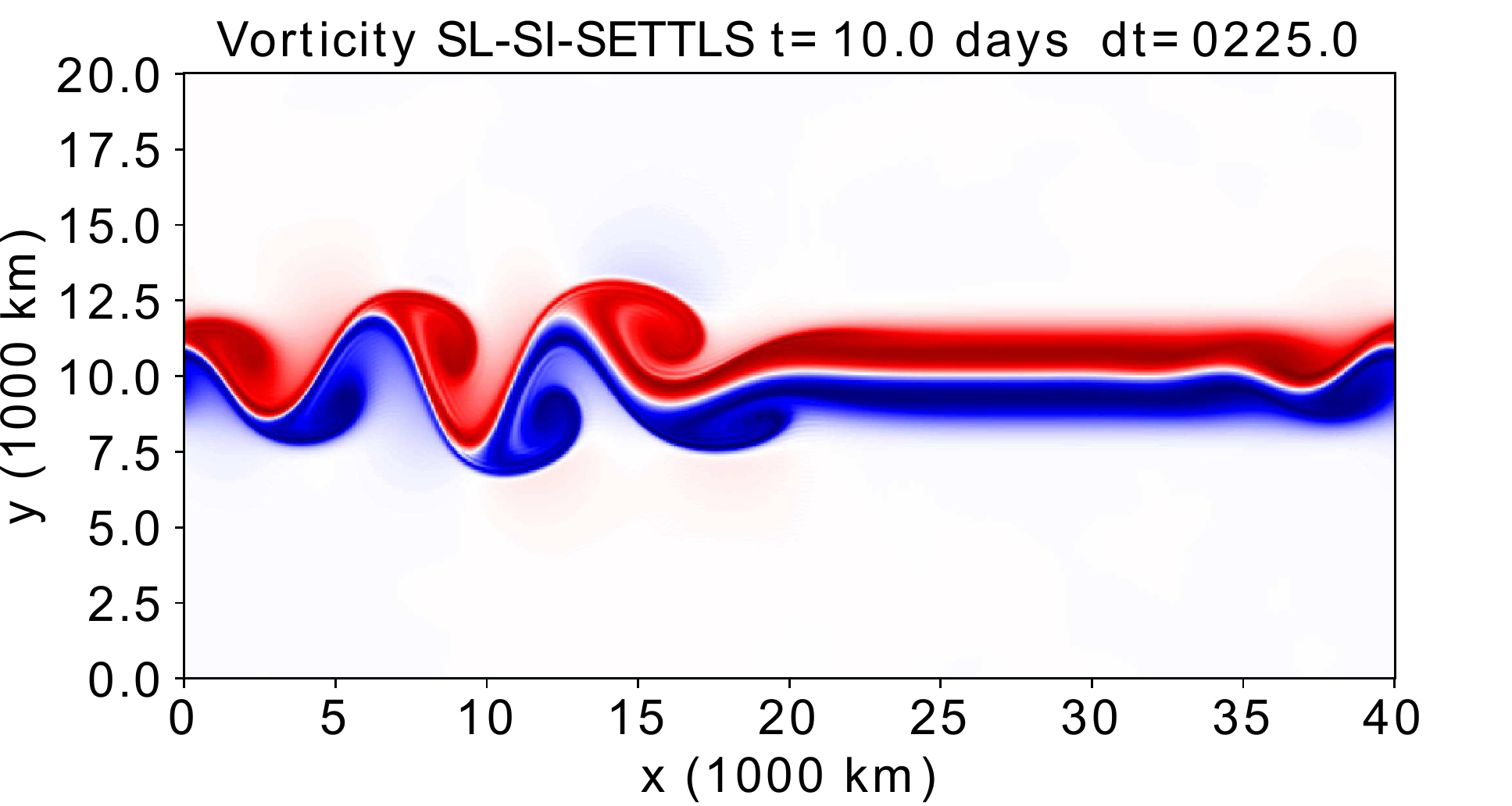}
	\zoomedareaimage{figures/{unstablejetscript_sl-rexi_g9.80616_h10000_f0.00014584_u0.0_U0_tsm_l_cn_na_sl_nd_settls_tso2_tsob2_C0225.0_REXIDIR_M0512_diag_vort_t00000864000.00000000halfcut}.pdf}	
	\includegraphics[scale=0.3]{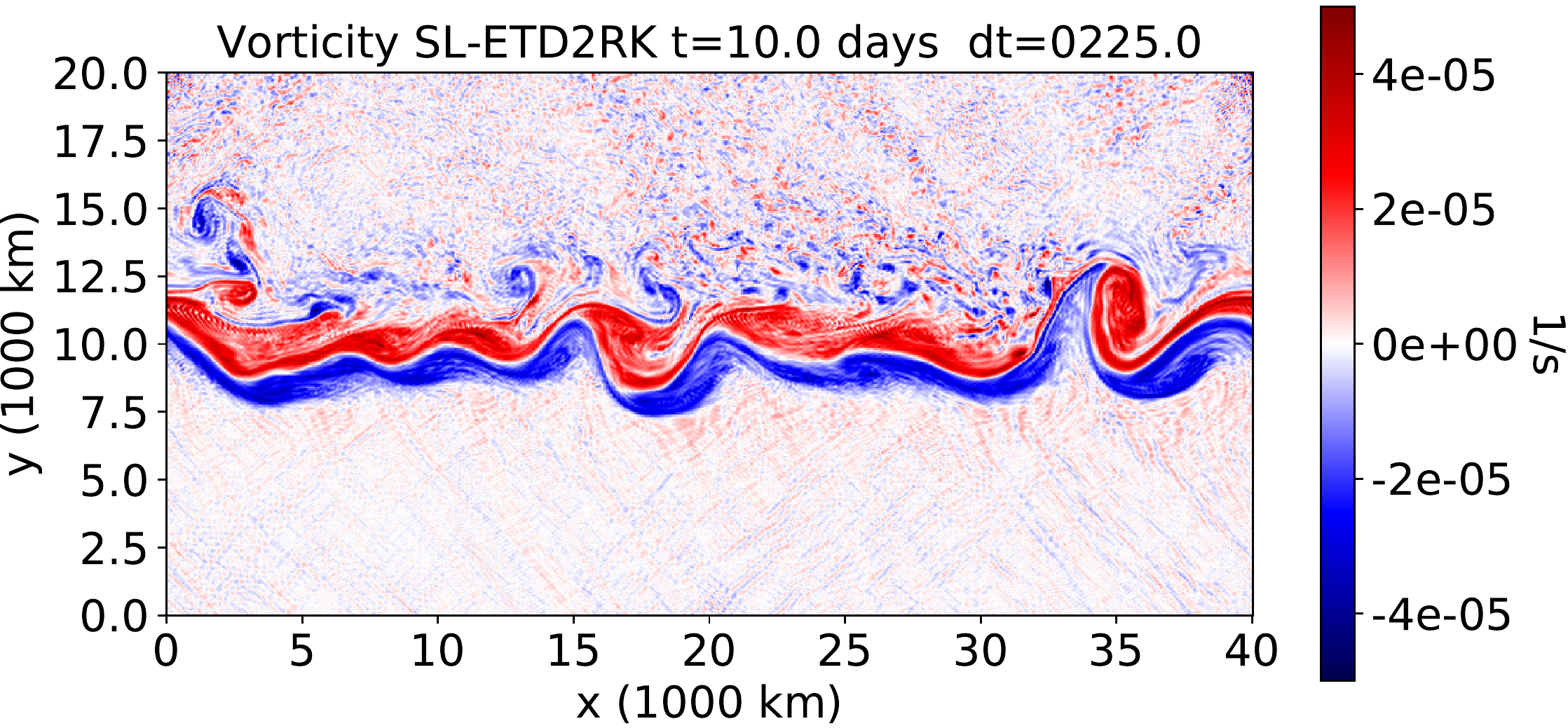}
	\zoomedareaimagex{figures/{unstablejetscript_sl-rexi_g9.80616_h10000_f0.00014584_u0.0_U0_tsm_l_rexi_na_sl_nd_etdrk_tso2_tsob2_C0225.0_REXIDIR_M0512_diag_vort_t00000864000.00000000half}.pdf}\\
	\begin{flushleft}
		\vspace{-0.7cm}
		(a) \hspace{5cm} (b)
	\end{flushleft}
	\caption{Numerical solution of the full nonlinear SWE at time 10 days for the vorticity field using a time-step size of 225 seconds. (a) SL-SI-SETTLS, (b) SL-ETD2RK.}
	\label{fig:fullswe_vorticity225sec}
\end{figure}

For larger time-step sizes, due to the extra energy in the high wavenumber gravity waves, the SL-ETD2RK triggers small turbulent like features after long runs when compared to SL-SI-SETTLS which is illustrated in Figure \ref{fig:fullswe_vorticity225sec}.

\begin{figure}[!ht]
	\centering
	\includegraphics[scale=0.4]{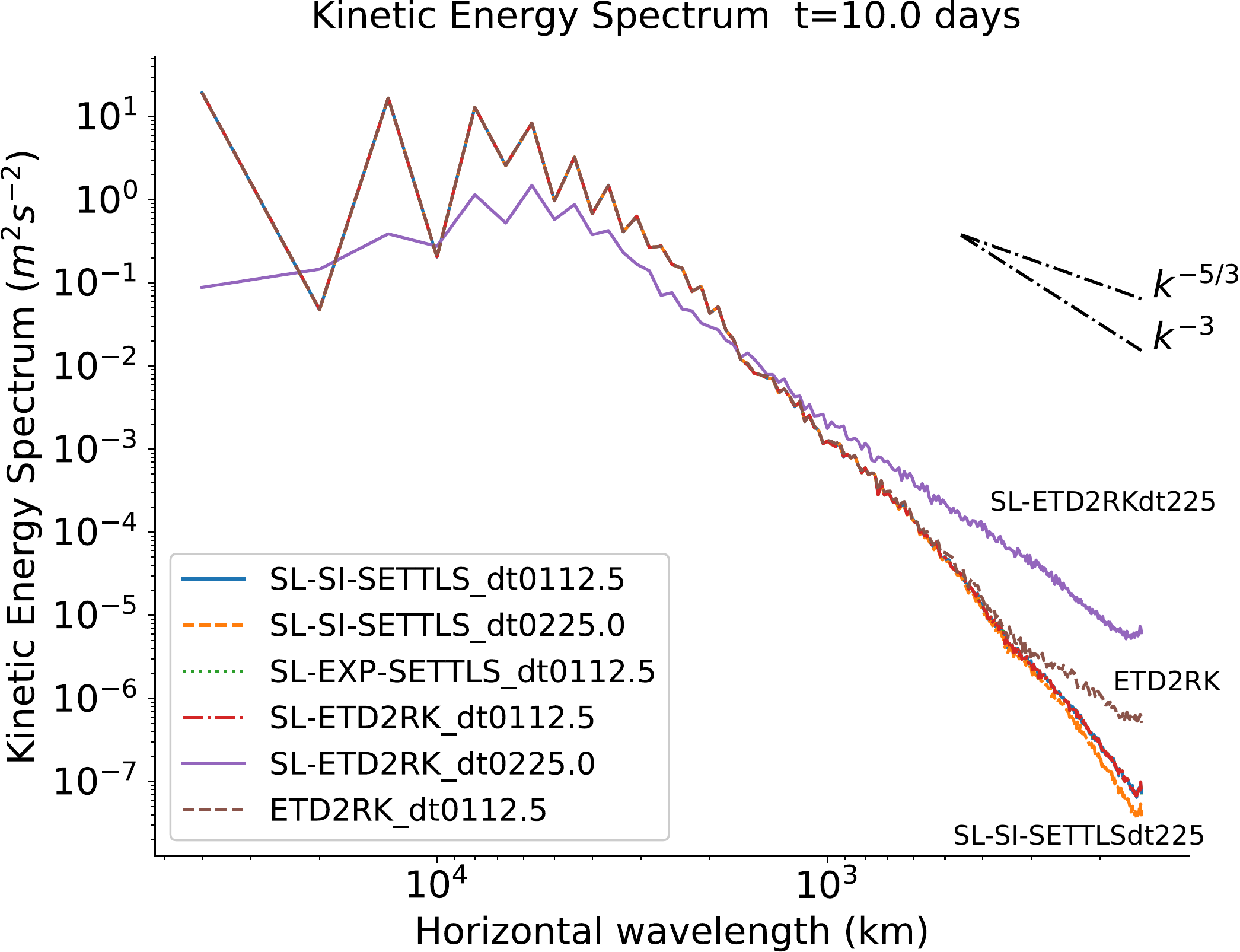}
	\caption{Kinetic energy spectrum for different methods and time-step sizes for the full nonlinear SWE at day 10 of integration.
		We can observe that the scheme SL-ETD2RK with 225 seconds time-step has developed less energy in large wavelengths and more energy in small wavelengths. The ETD2RK scheme has more energy in the small wavelengths and SL-SI-SETTLS with time-step of 225 seconds has the least amount of energy in small scales.
	}
	\label{fig:destabalized_ke_spec}
\end{figure}

Finally, this motivates to investigate the energy spectrum for selected time integration methods and parameters, see Figure \ref{fig:destabalized_ke_spec}.
If there is no dissipation of near grid scale energy for the SL-ETD2RK scheme, this energy destabilizes the jet into smaller scale features.
This can be clearly observed in the energy spectrum, where we also notice that the ETD2RK scheme gathers more energy in the smaller scales and the SL-SI-SETTLS with large time-step has the least amount of energy in the smaller scales.


\subsection{Shallow Water Equations with term specific diffusion}
For the purpose of weather and climate simulations, a certain amount of small-scale dissipation is usually required, either from a numerical stability perspective or from a physical point of view. The SL-SI-SETTLS scheme, when used in the full IFS dynamical core, adopts a spectral hyper-diffusion filter in the momentum equations in order to both numerically stabilize the scheme and physically dissipate energy from the small-scale energy tail (see \cite{gelb2001spectral} for an analysis of the impacts of the diffusion in a global spectral model and \cite{lauritzen2011numerical} for a comprehensive discussion on the use of diffusion in atmospheric models).
We remark that in full models this  
energy in high wavenumbers could be used to model physical sub-grid properties, such as convection.

With the semi-Lagrangian exponential scheme, it is possible to preserve the precise dispersion of linear waves and apply a term specific dissipation in the nonlinear divergence term.
This way, linear waves (long and short) are treated accurately, but only the longer waves originated from the nonlinear interactions are preserved in the model. This allows the model to be numerically stable without damping the linear waves, and also provides dissipation of small-scale features generated by the additional energy in high wavenumbers excited by the exponential integration.

\begin{figure}[!ht]
	\centering
	\includegraphics[scale=0.3]{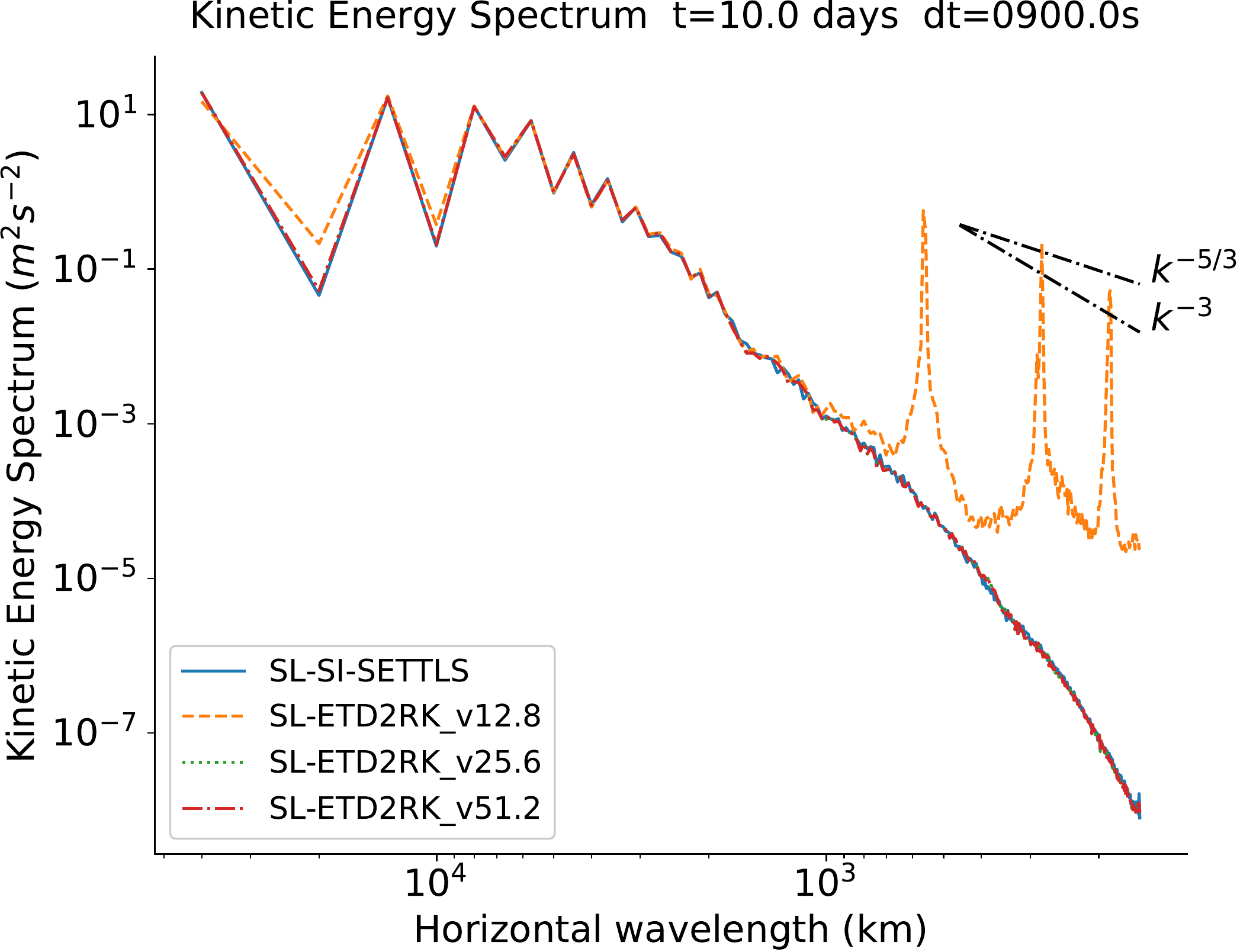}
	\includegraphics[scale=0.3]{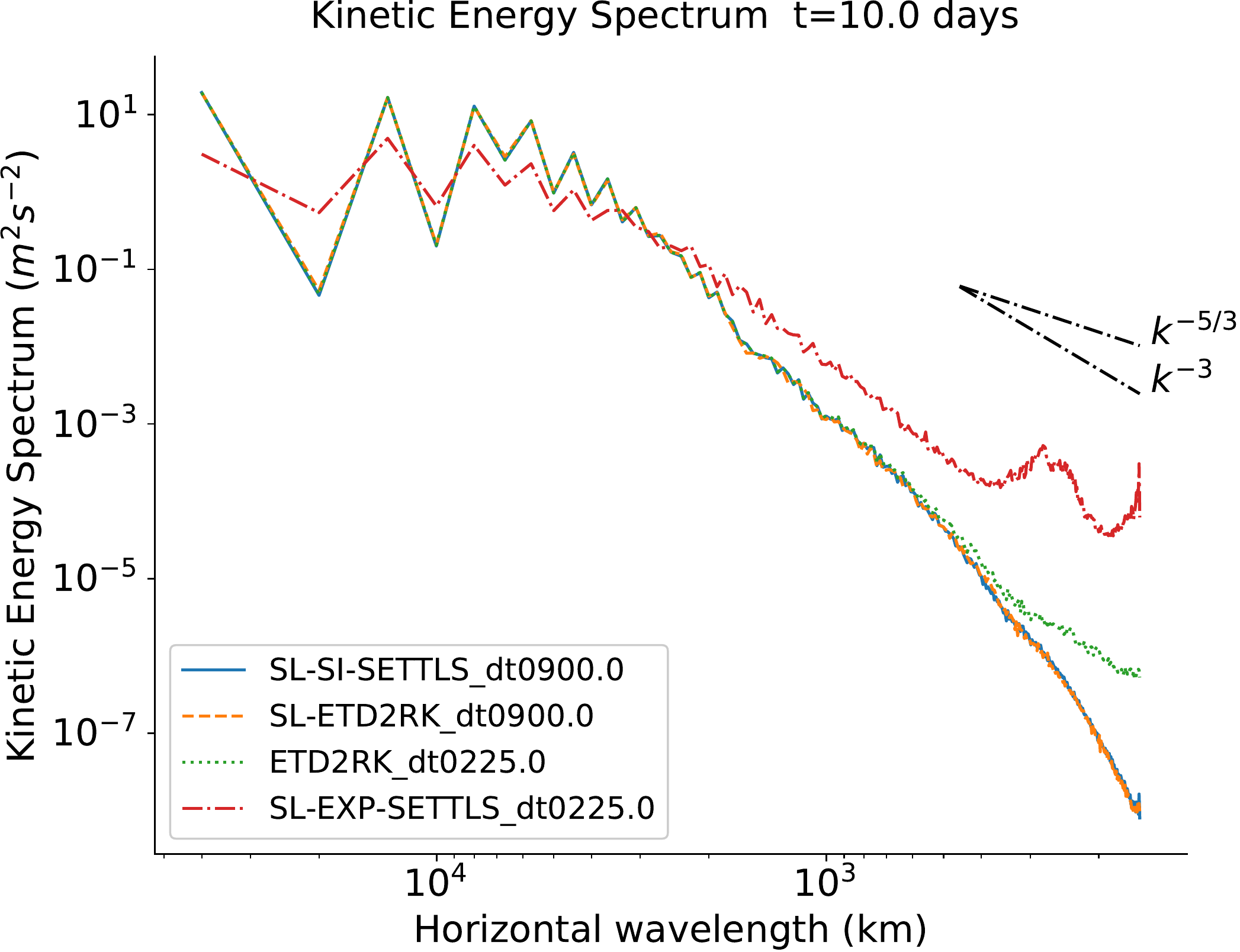}
	\begin{flushleft}
		\vspace{-0.5cm}
		(a) \hspace{6cm} (b)
	\end{flushleft}
	\caption{(a) Kinetic energy spectrum considering an implicit diffusion filter on the nonlinear divergence term various values of  $\mu$ ($12.8, \, 25.6 $ and $51.2\times 10^6 \,\mathrm{m^2s^{-1}}$), for the SL-ETD2RK scheme and no diffusion for the SL-SI-SETTLS scheme.
		(b) Kinetic energy spectrum considering an implicit diffusion on the nonlinear divergence term with $\mu=25.6\times 10^6 \,\mathrm{m^2s^{-1}}$, for the schemes and parameters shown in Figure \ref{fig:fullswe_vorticity225sec_visc256}.}
	\label{fig:kespec_full_swe_conv_dif}
\end{figure}

In the analysis that follows we considered an implicit spectral diffusion filter ($\mu \nabla^2 $) applied only to the nonlinear divergence term with $\mu$ the diffusion coefficient.

\begin{figure}[!ht]
	\centering
	(a)\includegraphics[scale=0.25]{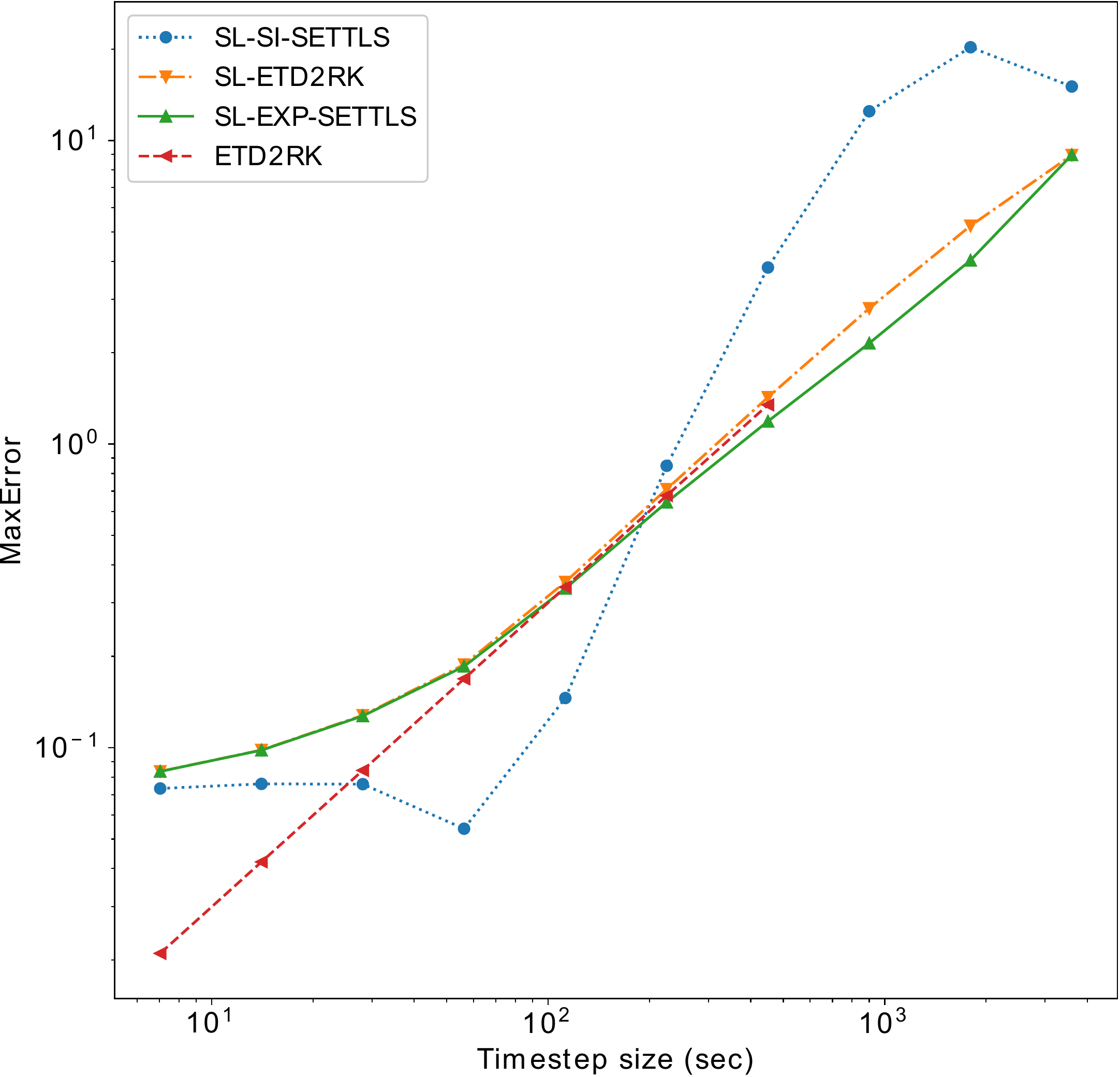}
	(b) \includegraphics[scale=0.25]{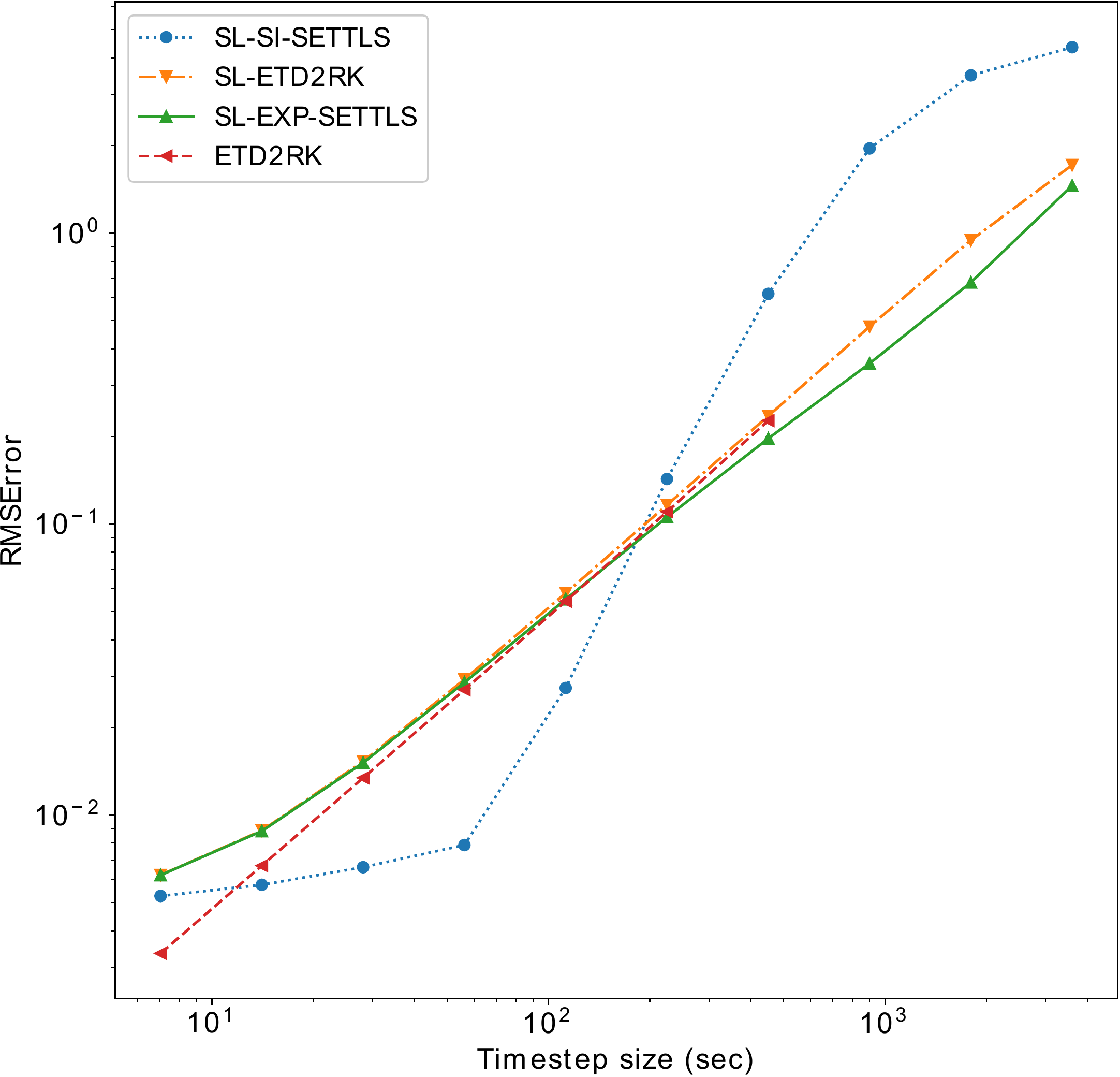}
	\hspace{0.5cm}
	\caption{Errors at day 1 of integration of the
		unstable jet test case for the full nonlinear SWE with implicit diffusion on the nonlinear divergence term ($\mu=25.6\times 10^{6} \,\mathrm{m^2s^{-1}}$). }
	\label{fig:full_swe_conv_dif}
\end{figure}

%
%
%
We start by analyzing different diffusion coefficients with the kinetic energy spectrum of the SL-ETD2RK scheme at day 10.
Figure \ref{fig:kespec_full_swe_conv_dif}a shows the amount of diffusion required to obtain a solution along the lines of the SL-SI-SETTLS with a time-step size of 900 seconds, and, following these results, we will adopt $\mu=25.6\times 10^{6} \,\mathrm{m^2s^{-1}}$. This value is similar to what is actually used in weather forecasting systems for the full equations, whereas here, we are only considering it for the nonlinear divergence.

Next, we investigate error comparisons at day 1 of integration shown in Figure\,\ref{fig:full_swe_conv_dif}.
We can observe that the two semi-Lagrangian exponential schemes deliver more accuracy compared to the SL-SI-SETTLS scheme.
However, the following results will reveal significant differences in the two semi-Lagrangian exponential schemes for larger time integration ranges.

\begin{figure}[!ht]
	\flushleft
	\includegraphics[scale=0.3]{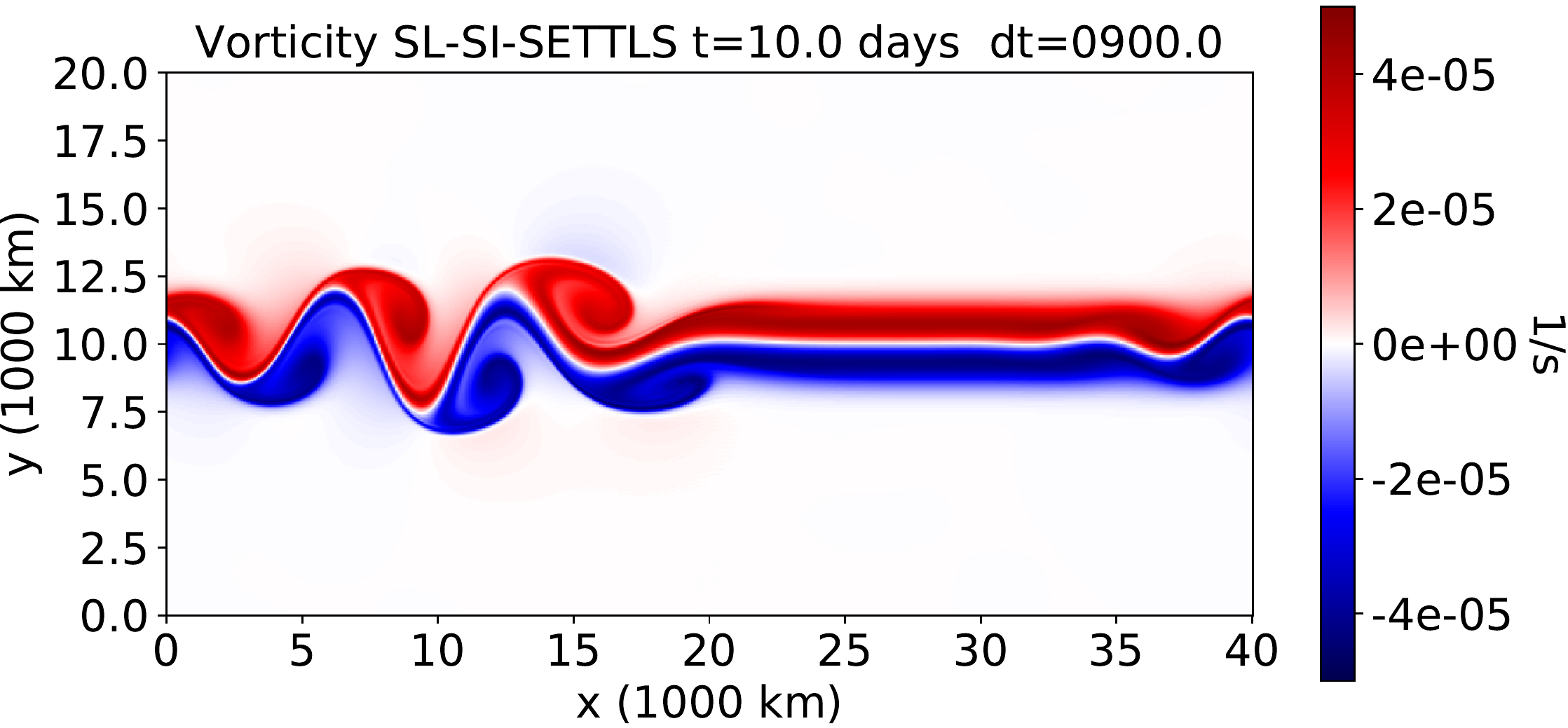}
	\zoomedareaimagea{figures/{script_sl-rexi_g9.80616_h10000_f0.00014584_u25600000_U2_tsm_l_cn_na_sl_nd_settls_tso2_tsob2_C0900.0_REXIDIR_M0512_diag_vort_t00000864000.00000000half}.pdf}
	\includegraphics[scale=0.3]{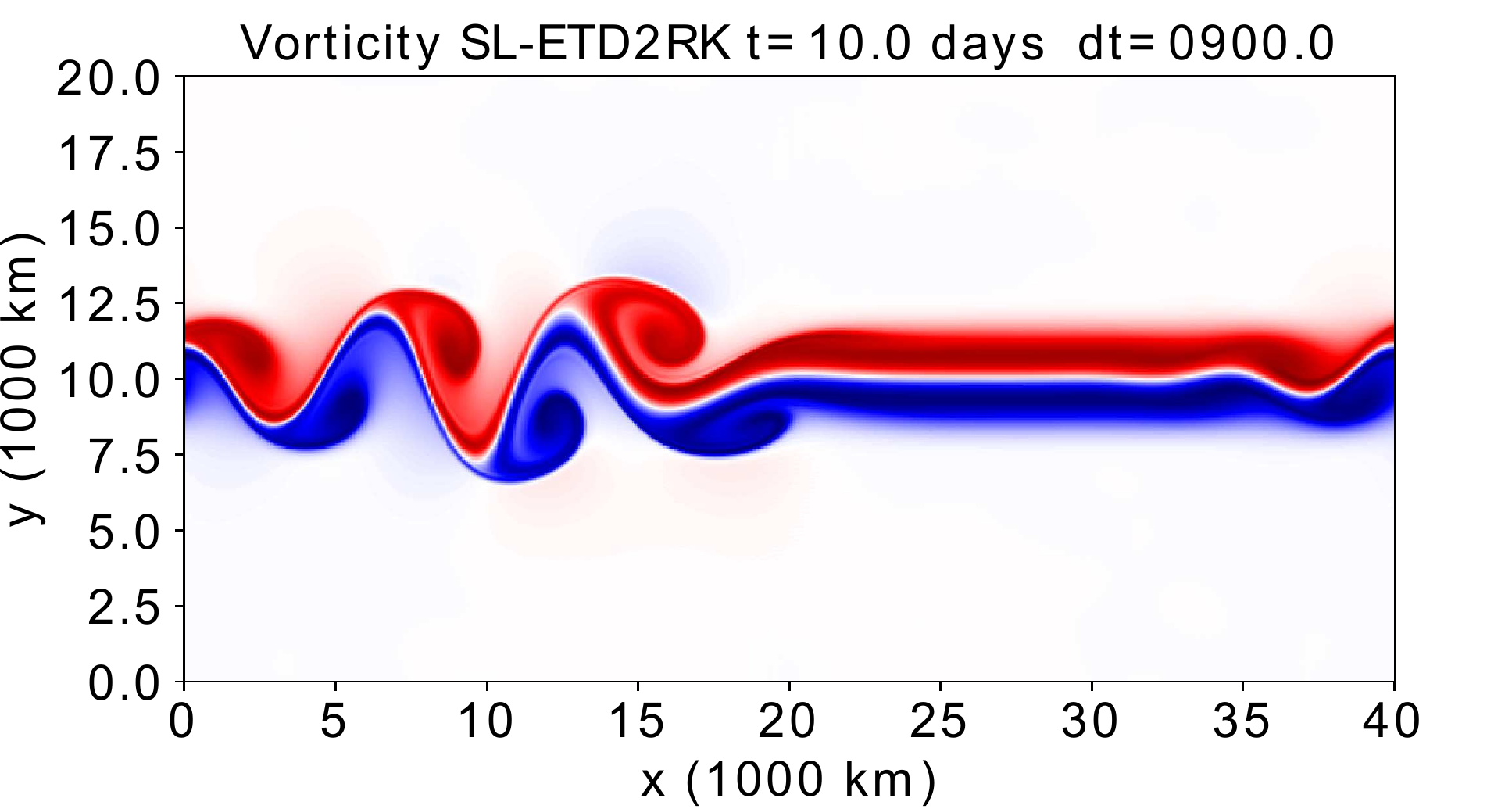}\
	\zoomedareaimageax{figures/{script_sl-rexi_g9.80616_h10000_f0.00014584_u25600000_U2_tsm_l_rexi_na_sl_nd_etdrk_tso2_tsob2_C0900.0_REXIDIR_M0512_diag_vort_t00000864000.00000000halfcut}.pdf}\\
	\vspace{-0.5cm}
	\flushleft (a) \hspace{6cm} (b) \\
	\includegraphics[scale=0.3]{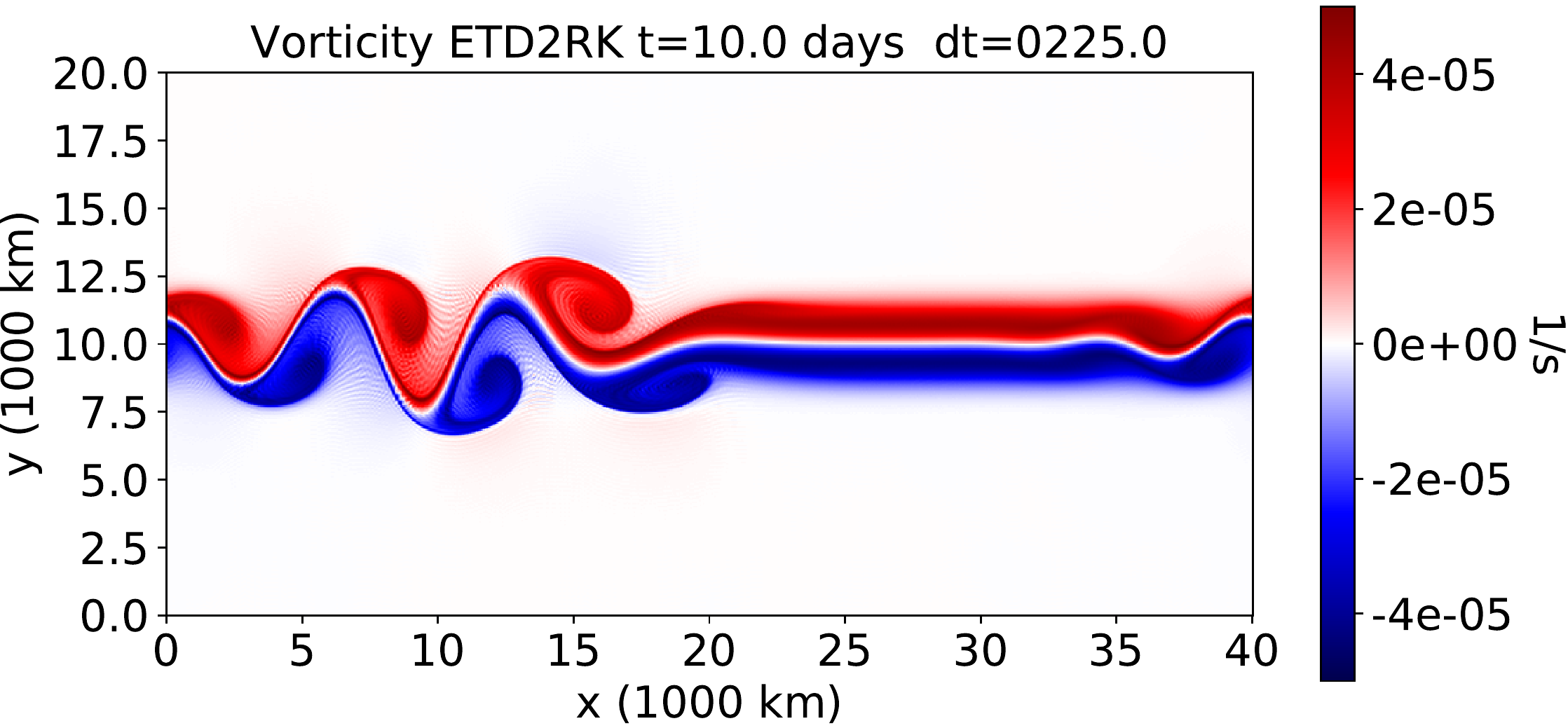}
	\zoomedareaimagea{figures/{script_sl-rexi_g9.80616_h10000_f0.00014584_u25600000_U2_tsm_l_rexi_n_etdrk_tso2_tsob2_C0225.0_REXIDIR_M0512_diag_vort_t00000864000.00000000half}.pdf}
	\includegraphics[scale=0.3]{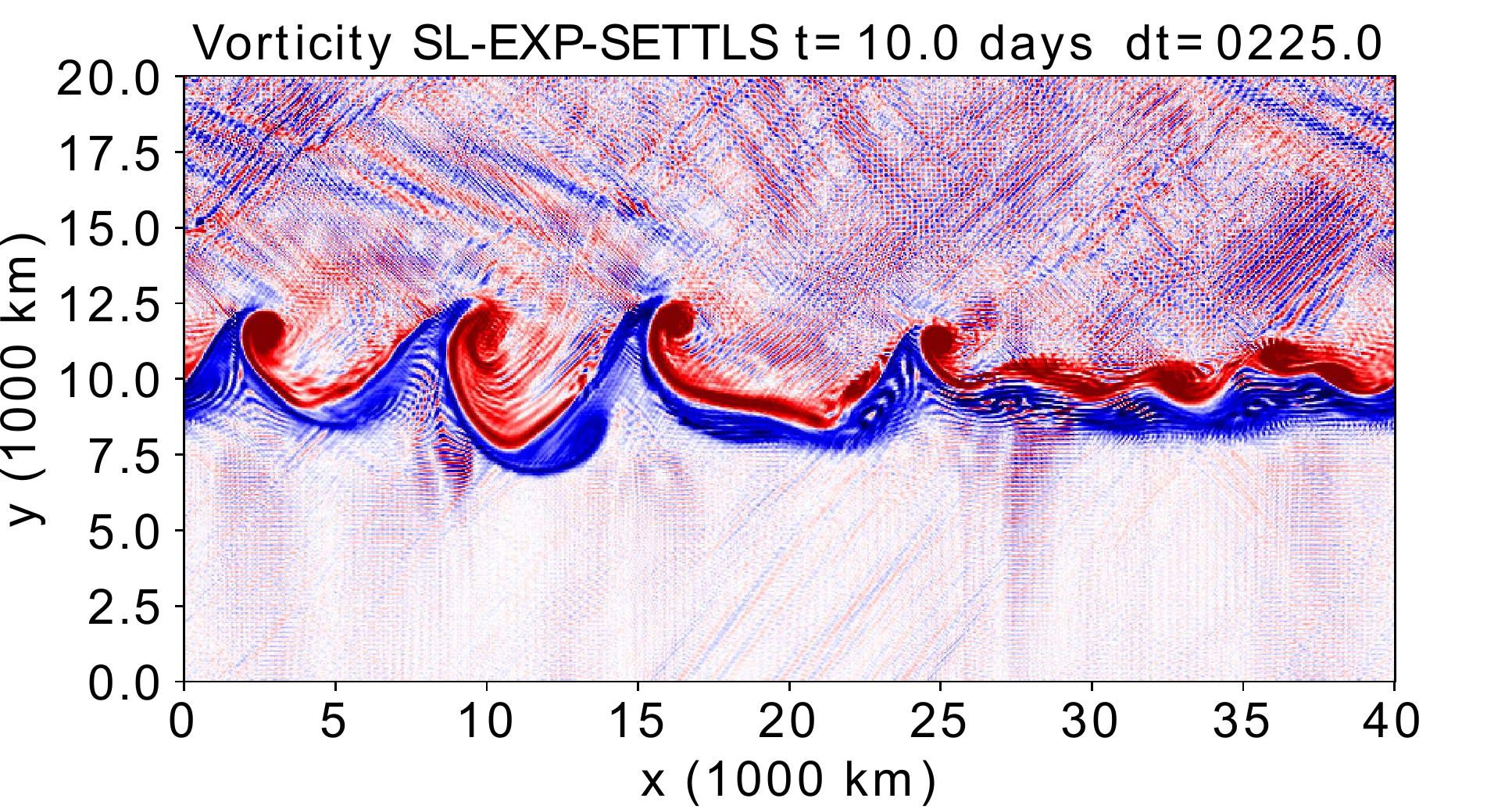}
	\zoomedareaimageax{figures/{script_sl-rexi_g9.80616_h10000_f0.00014584_u25600000_U2_tsm_l_rexi_na_sl_nd_settls_tso2_tsob2_C0225.0_REXIDIR_M0512_diag_vort_t00000864000.00000000halfcut}.pdf}\\
	\vspace{-0.5cm}
	\flushleft (c) \hspace{6cm} (d) \\
	\caption{Numerical solution of the full SWE at time 10 days for the vorticity field using implicit diffusion on the nonlinear divergence term with $\mu=25.6\times 10^{6} \,\mathrm{m^2s^{-1}}$. (a) SL-SI-SETTLS with $\Delta t = 900 s$, (b) SL-ETD2RK with $\Delta t = 900 s$, (c) ETD2RK with $\Delta t = 225 s$, (d) SL-EXP-SETTLS with $\Delta t = 225 s$.}
	\label{fig:fullswe_vorticity225sec_visc256}
\end{figure}

Finally, we time integrate to day 10 and investigate the different time integration methods, with the vorticity field depicted in Figure \ref{fig:fullswe_vorticity225sec_visc256}.
Despite the implicit diffusion, the ETD2RK scheme is still not able to do time-step sizes as large as the semi-Lagrangian schemes, due to the instability originated from the nonlinear advection term.
In Figure \ref{fig:kespec_full_swe_conv_dif} we show the kinetic energy spectra of the results shown in \ref{fig:fullswe_vorticity225sec_visc256}, where one may note the effects of the filtering in the small scale features. 
As before, the SL-EXP-SETTLS scheme is unstable for large time-steps and the diffusion was not able to circumvent this instability.
However, the SL-ETD2RK scheme with diffusion now does not develop near grid scale features (see Fig.\,\ref{fig:fullswe_vorticity112sec}b for comparison) even with a time-step size of 900 seconds.

\section{Concluding remarks}
\label{sec:conc}

Semi-Lagrangian schemes and exponential integrators both play important roles in different applications. The exploration of a mixed formulation is a challenging problem, partially tackled in this paper. We show a novel approach that combines these methods by exploring the exponential integration formulation in terms of material derivatives. 

The approach may be helpful for users of standard exponential integration techniques, extending these to larger time-step sizes preserving accurate solutions. 
Also, from the application of weather and climate modeling perspective, the method shows a way to improve the dispersion properties of a well-established scheme, therefore better representing fast linear gravity waves, with competitive computational cost.

The results presented in this paper show the potential benefits of the combination of these different methods, semi-Lagrangian and exponential integration, in a specific scenario, of rotating shallow water equations. Nevertheless, it also points out means of using such an approach in other equation sets that present both nonlinear advection and stiff linear operators, such as the Euler equations in fluid dynamics. We highlight, however, that the development of exponential schemes  for fluids that develop shocks, for which the use of conserved variables are usually preferred, still represent a challenge.





\section*{Acknowledgements}
The ideas behind this work originated from discussions with Colin Cotter, Jemma Shipton and Beth Wingate, whom are greatly acknowledged. We would also like to thank Saulo Barros for discussions with respect to semi-Lagrangian spectral schemes.

\appendix


\section{Lack of commutation between linear operator and interpolation at trajectory points}
\label{ap:slexpprop}
Consider a general vector $\vec{w} \in \mathbb{R}^n$, a linear operator 
$T \in \mathbb{R}^n \times \mathbb{R}^n$,  which will represent here, for example, a matrix exponential, and $\mathcal{I}_{\vec{x}}:\mathbb{R}^n \rightarrow \mathbb{R}^n$ an interpolation operation with respect to points $\vec{x} \in \mathbb{R}^n$. Following the semi-Lagrangian notation for interpolation, we may concisely write that $\mathcal{I}_{\vec{x}}(\vec{w})=\vec{w}_*$, where the $_*$ implicitly indicates the interpolation with respect to $\vec{x}$. This section is just to point a simple example to illustrate that even in very simple cases $(T\vec{w})_*\neq T(\vec{w}_*)$. 

Consider a 1D periodic grid with uniformly spaced points $(x_i)_{i=1,n}$, with distance $\Delta x$ from each other. In this example we will consider a scalar advection with constant velocity given by $\Delta x/ \Delta t$, so that, after one time-step, the departure points will be a simple translation and will match exactly their left neighbours. That is, the trajectory goes from $t_n$ to $t_{n+1}$ carrying the function value at $x_{i-1}$ to the $ x_{i}$ point. In this case, the interpolation to departure points will be given by a periodic shift in the indexes,
\begin{equation}
\mathcal{I}_{\vec{x}}(\vec{w}) = \mathcal{I}_{\vec{x}}([w_1,w_2,w_3,\dots,w_n])=[w_n,w_1,w_2,\dots,w_{n-1}]=\vec{w}_*.
\end{equation}
Note that the operator $\mathcal{I}_{\vec{x}}$ is a linear operator.

Now consider a simple diagonal linear operator $T=(\alpha_{ii})_{i=1,n}$, with $\alpha_{ii}\neq\alpha_{jj}$, for $j\neq i$. 
In this case,
\begin{eqnarray}
(T\vec{w})_*&=&([\alpha_{11}w_1,\alpha_{22}w_2,w_3,\dots,\alpha_{nn}w_n] )_* \\
\nonumber &=&[\alpha_{nn}w_n, \alpha_{11}w_1,\alpha_{22}w_2,w_3,\dots,\alpha_{(n-1)(n-1)}w_{n-1}],
\end{eqnarray}
but 
\begin{equation}
T(\vec{w}_*)=T[w_n,w_1,w_2,\dots,w_{n-1}]=[\alpha_{11}w_n,\alpha_{22}w_1,\dots,\alpha_{nn}w_{n-1}].
\end{equation}
Therefore, even if the trajectories are constant (or linear), the commutation does not generally hold.

In the more general case treated in the derivation of the semi-Lagrangian exponential scheme, the trajectories are nonlinear. Also, the linear operator is not necessarily diagonal, but one could think of its diagonalized version in complex space in a similar way, for which the terms in the diagonal would be the eigenvalues of the operator.

\bibliographystyle{siamplain}
\bibliography{bibliography}

\end{document}